\documentclass{aa}
\usepackage[varg]{txfonts}
\usepackage{graphicx,amssymb,epsfig,color,natbib}
\usepackage{times,graphics,amssymb,epsfig,subfigure,color}

\newcommand{\ber}{\begin{eqnarray}}
\newcommand{\eer}{\end{eqnarray}}

\def\apj{ApJ}

\def\aap{A\&A}

\def\labequn #1{\label{eq:#1}}
\def\labfig #1{\label{fig:#1}}
\def\labsecn #1{\label{sec:#1}}
\def\labsubsecn #1{\label{subsecn:#1}}
\def\labsubsubsecn #1{\label{subsubsecn:#1}}

\def\equn #1{Equation~\ref{eq:#1}}
\def\dequn#1#2{Equations~{\ref{eq:#1}}~and~{\ref{eq:#2}}}
\def\fig #1{Figure~\ref{fig:#1}}
\def\dfig #1#2{Figures~{\ref{fig:#1}}~and~{\ref{fig:#2}}}
\def\secn #1{Section~\ref{sec:#1}}
\def\dsecn #1#2{Sections~{\ref{sec:#1}}~and~{\ref{sec:#2}}}
\def\subsecn #1{Section~\ref{subsecn:#1}}
\def\subsubsecn #1{Section~\ref{subsubsecn:#1}}

\def\etal{et al.\ }

\def\unit #1{\,{\rm #1}}

\def\kev{\unit{keV}}

\bibpunct{(}{)}{;}{a}{}{,}

\begin{document}

\title{Absorption lines from magnetically-driven winds in X-ray binaries}

\author{S. Chakravorty\inst{1,2} \and P-O. Petrucci\inst{1,2} \and J. Ferreira\inst{1,2} \and G. Henri\inst{1,2} \and R. Belmont\inst{3,4} \and M. Clavel\inst{5} \and S. Corbel\inst{5} \and J. Rodriguez\inst{5} \and M. Coriat\inst{3,4} \and S. Drappeau\inst{3,4} \and J. Malzac\inst{3,4}}
\institute{Univ. Grenoble Alpes, IPAG, F-38000 Grenoble, France\\e-mail: susmita.chakravorty@obs.ujf-grenoble.fr \and CNRS, IPAG, F-38000 Grenoble, France \and Universit\'e de Toulouse; UPS-OMP; IRAP, F-31028, Toulouse, France \and CNRS; IRAP; 9 Av. colonel Roche, F-31028, Toulouse, France \and Laboratoire AIM (CEA/IRFU - CNRS/INSU - Universit\'e Paris Diderot), CEA DSM/IRFU/SAp, F-91191 Gif-sur-Yvette, France
}

\date{}
%%%%%%%%%%%%%%%%%%%%%%%%%%%%%%%%%%%%%%%%%%%%%%%%%%%%%%%%

\abstract
{High resolution X-ray spectra of black hole X-ray binaries (BHBs) show
blueshifted absorption lines suggesting presence of outflowing winds. Further,
observations show that the disk winds are equatorial and they occur in the
Softer (disk-dominated) states of the outburst and are less prominent or absent
in the Harder (power-law dominated) states.}
{We want to test if the self-similar magneto-hydrodynamic (MHD)
accretion-ejection models can explain the observational results for accretion
disk winds in BHBs. In our models, the density at the base of the outflow from
the accretion disk is not a free parameter. This mass loading is determined by
solving the full set of dynamical MHD equations without neglecting any physical
term. Thus the physical properties of the outflow depend on and are controlled
by the global structure of the disk.}
{We studied different MHD solutions characterized by different values of (a)
the disk aspect ratio ($\varepsilon$) and (b) the ejection efficiency
($p$). We also generate two kinds of MHD solutions depending on the absence
(cold solution) or presence (warm solution) of heating at the disk surface.
Such heating could be from e.g. dissipation of energy due to MHD turbulence in
the disk or illumination. Warm solutions can have large ($ > 0.1$) values of
$p$ which would imply larger wind mass loading at the base of the outflow. We
use each of these MHD solutions to predict the physical parameters (e.g.
distance density, velocity, magnetic field etc.) of an outflow. We have put
limits on the ionization parameter ($\xi$), column density and timescales,
motivated by observational results. Further constraints were derived for the
allowed values of $\xi$ from thermodynamic instability considerations,
particularly for the Hard SED. These physical constraints were imposed on each
of these outflows to select regions within it, which are consistent with the
observed winds.}
{The cold MHD solutions are found to be inadequate to account for winds due to
their low ejection efficiency. On the contrary warm solutions can have
sufficiently high values of $p (\gtrsim 0.1)$ which is required to explain the
observed physical quantities in the wind.  From our thermodynamic equilibrium
curve analysis for the outflowing gas, we found that in the Hard state a range
of $\xi$ is unstable. This constrain makes it impossible to have any wind at
all, in the Hard state.} 
{Using the MHD outflow models we could explain the observed trends - that the
winds are equatorial and that they are observable in the Soft states (and not
expected in the Hard state) of the BHB outbursts. }

%%%%%%%%%%%%%%%%%%%%%%%%%%%%%%%%%%%%%%%%%%%%%%%%%%%%%%%%%%%%%%%%%%%


%%%%%%%%%%%%%%%%%%%%%%%%%%%%%%%%%%%%%%%%%%%%%%%%%%%%%%%%%%%%%%%%%%%%

\maketitle

%%%%%%%%%%%%%%%%%%%%%%%%%%%%%%%%%%%%%%%%%%%%%%%%%%%%%%%%%%%%%%%%%%%%
\section{Introduction}
\labsecn{sec:introduction}

The launch of \textit{Chandra} and XMM-Newton, revealed blueshifted absorption
lines in the high resolution X-ray spectra of stellar mass black holes in
binaries (BHBs).  These are signatures of winds from the accretion disk around
the black hole. The velocity and ionization state of the gas, interpreted from
the absorption lines, vary from object to object and from observation to
observation. In most cases, only H- and He-like Fe ions are detected (e.g
\citealt{lee02}, \citealt{neilsen09} for GRS 1915+105, \citealt{miller04} for
GX 339-4, \citealt{miller06} for H1743-322 and \citealt{king12} for IGR
J17091-3624). In some of the objects however, a wider range of ions is seen
from O through Fe (e.g. \citealt{ueda09} for GRS 1915+105, \citealt{miller08,
kallman09} for GRO J1655--40). The variations in the wind properties seem to
indicate variations in the temperature, pressure and density of the gas from one
object to another. Further, even in the same object, the winds seem to have
variations depending on the accretion state of the black hole.

Both spectral and timing observations of most BHBs show common behaviour
patterns centered around a few states of accretion. The spectral energy
distributions (SEDs) corresponding to the different states have varying degree
of contribution from the accretion disk and the non-thermal power-law
components. The X-ray studies of BHB show that winds are not present in all
states. It has been shown by several authors that the absorption lines are more
prominent in the Softer (accretion disk dominated) states \citep{miller08,
neilsen09, blum10, ponti12}. For some objects, the reason for such changes is
attributed to changes in the photoionizing flux \citep[e.g.][in the case of
H1743-322]{miller12}. However, the alternative explanation of `changes in the
driving mechanism' is of greater relevance to this paper. 

The observable properties of the accretion disk winds are often used to infer
the driving mechanism of the winds \citep{lee02, ueda09, ueda10, neilsen11,
neilsen12a}. Hence the variation or disappearance of the wind through the
various states of the BHB, has been interpreted as variation in the driving
mechanism of the wind. A good example is the case of GRO J1655--40. A well
known \textit{Chandra} observation of GRO J1655--40 \citep{miller06, miller08,
kallman09}, showed a rich absorption line spectrum from OVIII - NiXXVI, and led
the authors to conclude for magnetic driving mechanism for the wind.
\citet{neilsen12a} analysed the data from another observation from 3 weeks
later, for the same source, and found absorption by Fe\, {\sc xxvi} only. They
argue that such a change cannot be due to variation in photoionization flux
only and suggest that variable thermal pressure and magnetic fields may be
important in driving long-term changes in the wind in GRO J1655--40.

To get a consolidated picture of these systems, it is necessary to understand
the relation between the accretion states of the BHBs and the driving
mechanisms of the winds. In this paper we investigate the magneto hydrodynamic
(hereafter MHD) solutions as driving mechanisms for winds from the accretion
disks around BHBs - cold solutions from \citet[][hereafter F97]{ferreira97} and
warm solutions from \citet{casse00b} and \citet{ferreira04}. To understand the
basic motivation of the MHD solutions used to model the winds, throughout this
paper, it is important to discuss the distinction between winds and jets from
accretion disks. Observationally, jets are usually described as collimated,
fast (mildly relativistic) outflows detected or directly imaged in radio
wavelengths. On the other hand, winds are detected as absorption features,
showing speeds of a few thousand $\rm{km \, s^{-1}}$. However, on the
theoretical side, both are outflows launched from the accretion disk surface
due to magnetic and/or thermal/radiative effects. The power carried by these
outflows is, ultimately, a fraction of the released accretion power. Hence,
although observationally distinct, theoretically, it is not simple to
distinguish between the two. One way to make a clear theoretical distinction
between these two outflows is to look at the magnetization $\sigma$ at the disk
surface, namely the ratio of the MHD Poynting flux to the sum of the thermal
energy flux and the kinetic energy flux. Jets would have $\sigma > 1$, a high
magnetization translating into both large asymptotic speeds and (magnetic)
self-confinement. On the contrary, winds would be much less magnetized ($\sigma
<1$) with much lower asymptotic speeds and the confinement (if any) will come
only from the external medium. 

MHD solutions have been used by other authors to address outflows in various
systems. Of particular relevance to this paper are the works presented by
\citet{fukumura10a, fukumura10b, fukumura14, fukumura15}. Based on the
self-similar \citet{contopoulos94} MHD models of outflowing material, the
aforementioned papers have already argued in favour of large scale
magneto-centrifugally driven winds in active galactic nuclei (AGN - galaxies
which host actively mass accreting super-massive, $M_{BH} > 10^6 M_{\odot}$,
black holes at their centres). Their analysis shows that such models can
account for the observed warm absorbers and ultra-fast outflows seen as
absorption lines in high resolution X-ray spectra of AGN. They have also
attempted to explain the broad absorption lines (seen in high resolution
ultraviolet spectra of AGN) using the same MHD wind models. Note however, that
the \citet{contopoulos94} model (which is an extension of the
\citealt{blandford82} hydromagnetic flows ) does not treat the underlying disk.
As a consequence, the link between the mass loss in winds and the disk
accretion rate is lost and the mass loading at the base of the disk can be
(almost) arbitrarily large or small. On the contrary, the MHD models in F97
(and subsequent papers) link the density of the outflowing material to the disk
accretion rate. 

A consistent theory of MHD outflows from the disk must explain how much matter
from the disk is deviated from the radial to the vertical motion, as well as
the amount of energy and angular momentum carried away from the disk. This
requires a thorough treatment of the resistive disk interior and matching it
with the outflowing material using ideal MHD. The only way to solve such an
entangled problem is to take into account all dynamical terms, a task that has
been done within a self-similar framework in F97. 

The F97 MHD solutions have been used in \citet{ferreira06} and
\citet{petrucci10}, to describe accretion disks giving rise to jets in the Hard
States of BHBs. Winds, on the other hand are seen in the Soft state of the BHBs
when radio jets are absent. Using the F97 models we aim to test if the same
theoretical framework (which could reproduce jets) can reproduce the observed
properties of the winds (ionization parameter, column density, velocity etc.).
We shall further, look into the parameter space of the theoretical models to
distinguish between the Softer accretion states, when the wind is observed and
the Harder states when the absorption lines from the wind is not observed.

%%%%%%%%%%%%%%%%%%%%%%%%%%%%%%%%%%%%%%%%%%%%%%%%%%%%%%%%%%%%%%%%%%%%

%%%%%%%%%%%%%%%%%%%%%%%%%%%%%%%%%%%%%%

\section{The MHD accretion disk wind solutions}
\labsecn{sec:MhdWinds}

%%%%%%%%%%%%%%%%%%%%%%%%%%%%%
\subsection{General properties}
\labsubsecn{subsec:genprops}

We use the F97 solutions describing steady-state, axisymmetric solutions under
the following two conditions: \\
(1) A large  scale  magnetic  field  of  bipolar  topology  is  assumed to
thread the accretion disk. The strength of the required vertical magnetic field
component is obtained as a result of the solution \citep{ferreira95}.\\ 
(2) Some anomalous turbulent resistivity is at work, allowing the plasma to
diffuse through the field lines inside the disk.      
 
For a set of disk parameters, the solutions are computed from the disk midplane
to the asymptotic regime, the outflowing material becoming, first, super
slow-magnetosonic, then, Alfv\'enic and finally, fast-magnetosonic.  All
solutions that will be discussed in this paper, have this same asymptotic
behavior which corresponds to the following physical scenario: after an opening
of the radius of the outflow, leading to a very efficient acceleration of the
plasma, the outflow undergoes a refocusing towards the axis (recollimation).
The solutions are then, mathematically terminated (see F97 for more details).
Physically speaking however, the outflowing plasma will most probably undergo
an oblique shock (which is independent of the assumption concerning the thermal
state of the magnetic surfaces) after the recollimation happens. However,
theoretically accounting for the oblique shock is beyond the scope of this
paper. Thus, in this paper we rely on those solutions only, which cross their
Alfv\'en surfaces before recollimating (i.e. before the solutions have to be
mathematically terminated). 

%%%%%%%%%%%%%%%%%%%%%%%%%%%%%

%%%%%%%%%%%%%%%%%%%%%%%%%%%%%%%%%%%%%%%%%%%%%%%%%%%%%%%%%%%%%%%%%%%%%%%%%%%%%
\begin{figure}
\begin{center}
\includegraphics[width = 9 cm, angle = 0]{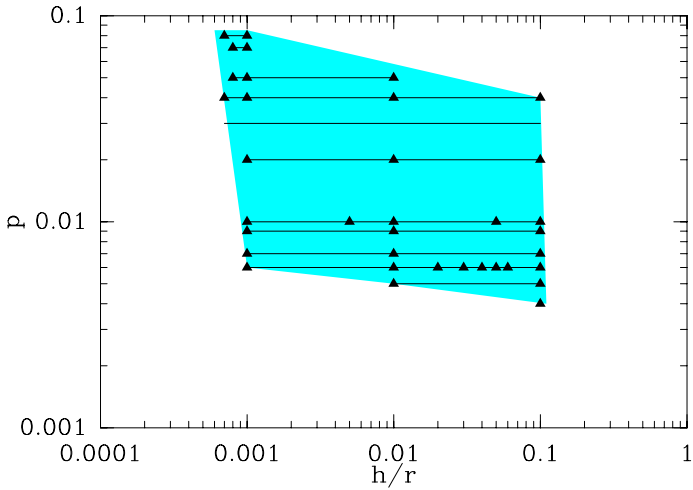}
\caption{Parameter space ejection index $p$ ($\dot M_{acc} \propto r^p$) versus
disk aspect ratio $\varepsilon= h/r$ for isothermal, cold accretion-ejection
solutions of F97. The colored area shows the zone where super-Alfvenic outflows
have been obtained, the triangles are some specific solutions and lines are for
constant $p$.} 
\labfig{fig:paramF97}
\end{center}
\end{figure}
%%%%%%%%%%%%%%%%%%%%%%%%%%%%%%%%%%%%%%%%%%%%%%%%%%%%%%%%%%%%%%%%%%%%%%%%%%%%%%

%%%%%%%%%%%%%%%%%%%%%%%%%%%%%
\subsection{Model parameters}
\labsubsecn{subsec:params}

The rigorous mathematical details of how the isothermal MHD solutions for the
accretion disk outflow are obtained are given in the aforementioned papers and
we refrain from repeating them here. In this section, we focus on describing
the two parameters that affect the density $n^+$ (or $\rho^+$) of the
outflowing material at a given radius $r$ in the disk. 

Because of ejection, the disk accretion rate varies with the radius even in a
steady state, namely $\dot M_{acc} \propto r^p$. This radial exponent, $p$
\citep[labelled $\xi$ in F97, ][etc.]{ferreira06, petrucci10} is very important
since it measures the local ejection efficiency.  For an accretion disk which
is giving rise to an outflow, the mass outflow rate is related to the accretion
rate through the ejection index p. If the disk extends between the inner radius
$r_{in}$ and the outer radius $r_{out}$ and is being fed by a disc accretion
rate $\dot M_{acc}(r_{out})$, the ejection to accretion mass rate ratio is 
\begin{equation}
\frac{2 \dot M_{outflow}}{\dot M_{acc}(r_{out})} = 1 - \left ( \frac{r_{in}}{r_{out}} \right )^p \simeq p \ln \frac{r_{out}}{r_{in}} 
\end{equation}
where the last estimate holds only for $p<<1$. For the most extreme MHD
solution discussed in this paper (namely $r_{out} \simeq 10^7 r_{in}$ and $p
\simeq 0.1$), about 80\% of the accreted mass is ejected in the form of the
outflow.  The larger the exponent, the more massive and slower is the outflow.
Mass conservation writes
\begin{eqnarray}
2 \frac{d \dot M_{outflow}}{dr}= 4 \pi r \rho^+ u^+_z &=& \frac{d \dot M_{acc}}{dr} = p \frac{\dot M_{acc}}{r} \nonumber \\
n^+ m_p = \rho^+ &\simeq & \frac{p}{\varepsilon} \frac{\dot M_{acc}}{4 \pi \Omega_K r^3} 
\labequn{eqn:rho+}
\end{eqnarray}
where $m_p$ is the proton mass and the superscript "+" stands for the height
where the flow velocity becomes sonic\footnotemark,
\footnotetext{Actually, the sonic speed only provides here a convenient scaling for the velocity, especially in isothermal flows. But in MHD winds, the critical speed that needs to be reached at the disk surface is the slow magnetosonic speed, which is always smaller than the sonic speed (see Appendix for more details).}
namely $u_z^+= C_s = \Omega_K h = \varepsilon V_K$. Here, $V_K=\Omega_K
r=\sqrt{GM_{BH}/r}$ ($G$: gravitational constant) is the keplerian speed and 
\begin{equation}
\varepsilon= \frac{h}{r}
\labequn{eqn:Epsilon}
\end{equation}
is the disk aspect ratio, where $h(r)$ is the vertical scale height at the
cylindrical radius $r$. It can thus be seen that the wind density, a crucial
quantity when studying absorption features, is mostly dependent on $p$ and
$\varepsilon$ for a given disk accretion rate $\dot M_{acc}$.  

\equn{eqn:rho+} is the fundamental difference between the MHD models used in
the aforementioned papers by Fukumura \etal and the ones used in this work.
While in the former, the initial wind density $\rho^+$ can be ``arbitrarily''
prescribed i.e. independent of the the underlying disk accretion rate, here it
is a result of an accretion-ejection calculation and are determined by $p$ and
$\varepsilon$. In the Fukumura \etal papers there are two assumptions, put by
hand, that determine the physical properties of the outflow. a) The authors do
not use the parameter $p$. However, comparing the equations for the radial
distributions of magnetic field ($B_z \propto r^{q-2}$) of the outflow, we can
get the relation  $q= \frac{3}{4} + \frac{p}{2}$ \citep{ferreira93}. Note that
$q$ is not any parameter related to the accretion disk, but an index related to
the outflow. The Fukumura \etal papers discuss the two cases of with $q=1$ and
$q=3/4$, but for modelling the AGN winds they use the former, which would
correspond to $p=0.5$. The choice of $q=1$ was to ensure that the density in
the outflow followed $n \propto r^{\alpha}$ with $\alpha = 2q - 3 = 1$, as
suggested by observations. b) The density at the launching point of the wind is
prescribed by a parameter $\eta_W$ which is the ratio of the mass outflow rate
to the disk accretion rate. Note that the authors use a constant value $\eta_W
= 0.5$, independent of $q$. These preassigned values for the parameters
defining the outflow and the lack of any connection to the accretion process,
fosters a sense of "physical arbitrariness". To achieve such a high value of
$\eta_W$, an extra process (other than magneto-hydrodynamic acceleration) must
be acting within the resistive disk (this will be discussed later in
\secn{sec:WarmSolns} in the context of ``Warm'' models).

In the MHD models used in this paper the value of the exponent $p$ influences
the extent of magnetisation in the outflow. This is another way in which the
ejection index relates the accretion process and the outflow properties. In a
non-relativistic framework the ratio of the MHD Poynting flux to the kinetic
energy flux at the disk surface is 
\begin{equation}
%\sigma^+ = \frac{| {\bf E} \times {\bf B} | }{\mu_o \rho u^2 u_p/2} \simeq
%\frac{1}{p} \left ( \frac{\Lambda}{1+ \Lambda} \right )
\sigma^+ \simeq \frac{1}{p} \left ( \frac{\Lambda}{1+ \Lambda} \right )
\labequn{eqn:mag_sigma}
\end{equation} \citep[F97,][]{casse00a} where $\Lambda$ is the ratio of the
torque due to the outflow to the turbulent torque (usually referred to as the
viscous torque).  The torque due to the outflow transfers the disk angular
momentum to the outflowing material whereas the turbulent torque provides an
outward radial transport within the disk. Smaller the value of $p$, larger is
the energy per unit mass in the outflow. A magnetically dominated self-confined
outflow requires $\sigma^+ >1$.  The F97 outflow models have been obtained in
the limit $\Lambda \rightarrow \infty$ so that  the self-confined outflows
carry away all the disk angular momentum and thereby rotational energy with
$\sigma^+ \simeq 1/p \gg 1$. The outflow material reaches the maximum
asymptotic poloidal speed $V_{max} \sim V_K(r_o) p^{-1/2}$, where $r_o$ is the
anchoring radius of the magnetic field line. 

\fig{fig:paramF97} shows the $p-\varepsilon$ parameter space of super-Alfvenic
MHD solutions obtained by F97 with cold, isothermal magnetic surfaces. It can
be seen that under these assumptions it is impossible to achieve high values of
$p \gtrsim 0.1$. 
%{\color{red} Renaud's comment: Is there a simple way to understand why
%(physical reasons) there is no solutions for large values of p?} 
Such a limit on the value of $p$ does not improve even if the magnetic surfaces
are changed to be adiabatic, as long as the outflowing material remains cold
\citep{casse00a}. The outflow is cold when its enthalpy is negligible when
compared to the magnetic energy, which is always verified in near Keplerian
accretion disks. However, the warming up of the outflowing material could occur
if some additional heat deposition becomes active at the disk surface layers
(through illumination for instance, or enhanced turbulent dissipation at the
base of the corona). 
In that case, larger values of $p$ up to $\sim 0.45$ have
been reported \citep{casse00b,ferreira04}. We will examine the cold outflows
in \secn{sec:EpPVar} and the ``warm outflows'' in \secn{sec:WarmSolns}.

%%%%%%%%%%%%%%%%%%%%%%%%%%%%%

%%%%%%%%%%%%%%%%%%%%%%%%%%%%%
\subsection{The scaling relationships}
\labsubsecn{subsec:scalings}

For the MHD outflow (with given $\varepsilon$ and $p$) emitted from the
accretion disk settled around a black hole, the important physical quantities
are given at any cylindrical (r,z) by 
\begin{equation}
n(r,z) = \frac{\dot m}{\sigma_T r_g} \left(\frac{r}{r_g}\right)^{(p-3/2)} f_n(y) 
%n(r) &\propto & \frac{\dot m}{\sigma_T r_g} \left(\frac{r}{r_g}\right)^{(p-3/2)} 
\labequn{eqn:n_scaling}
\end{equation}
\begin{equation}
B_i(r,z) = \left (\frac{\mu_o m_p \dot m c^2}{\sigma_T r_g}  \right )^{1/2}  \left(\frac{r}{r_g}\right)^{(-5/4 + p/2)} f_{B_i}(y)
\labequn{eqn:B_scaling}
\end{equation}
\begin{eqnarray}
v_i(r,z)  & = & c \left(\frac{r}{r_g}\right)^{-1/2} f_{v_i}(y) \\
\tau_{dyn}(r)  & = & \frac{2\pi r_g}{c} \left(\frac{r}{r_g}\right)^{3/2} f_{\tau}(y)
\end{eqnarray}
where $\sigma_T$ is the Thomson cross section, $c$ the speed of light, $r_g = G
M_{BH} / c^2$ is the gravitational radius, $\mu_o$ the vacuum magnetic
permeability, $y = z/r$ the self-similar variable and the functions $f_X(y)$
are provided by the solution of the full set of MHD equations. In the above
expressions, $n$ is the proton number density and we consider it to be $\sim
n_H$ (the Hydrogen number density); $v_i$ (or $B_i$) is any component of the
velocity (or magnetic field) and $\tau_{dyn} = 1/div \bf{V}$ (where $\bf{V}$ is
the plasma velocity) is a measure of the dynamical time in the flow. The
normalized disk accretion rate used in the above equations is defined by 
\begin{equation}
\dot m = \frac{\dot M_{acc}(r_g) \, c^2}{L_{Edd}} \nonumber
\labequn{eqn:mdot}
\end{equation}
where $L_{Edd}$ is the Eddington luminosity.

Note that we are using a steady state assumption for the accretion disk of a
BHB i.e. the variation of the accretion rate with the radius is assumed to be
the same for the entire disk (same $p$ and same
normalization). This assumption is maintained from the innermost regions (a few
$r_g$) to the outer part of the disk where the disk wind becomes relevant
(between $10^3 - 10^4 r_g$). We acknowledge that this is a simplistic picture
since BHBs are outbursting systems where the accretion rate is obviously
varying. So the accretion rate of the outer part of disk could be significantly
different to the one in the inner part. Taking this effect into account would
however, require considering a detailed time evolution of the accretion
mechanism through the different stages of the outburst which is far beyond the
scope of this paper. Hence we proceed forward to perform our calculations,
within the aforementioned scientific framework.

%%%%%%%%%%%%%%%%%%%%%%%%%%%%

%%%%%%%%%%%%%%%%%%%%%%%%%%%%%%%%%%%%%%%%%%%%%%%%%%%%%%%%%%%%%%%%%%%%%%%%%%%%%%

%%%%%%%%%%%%%%%%%%%%%%%%%%%%%%%%%%%%%%%%%%%%%%%%%%%%%%%%%%%%%%%%%%%%%%%%%%%%%%
\section{Observational constrains}
\labsecn{sec:ObsConstrns}

%%%%%%%%%%%%%%%%%%%%%%%%%%%%%%%%%%%%%%
\subsection{The spectral energy distribution for the Soft and the Hard state}
\labsubsecn{subsec:SED}

%%%%%%%%%%%%%%%%%%%%%%%%%%%%%%%%%%%%%%%%%%%%%%%%%%%%%%%%%%%%%%%%%%%%%%%%%%%%%
\begin{figure}
\begin{center}
\includegraphics[scale = 1, width = 9 cm, trim = 0 125 0 0, clip, angle = 0]{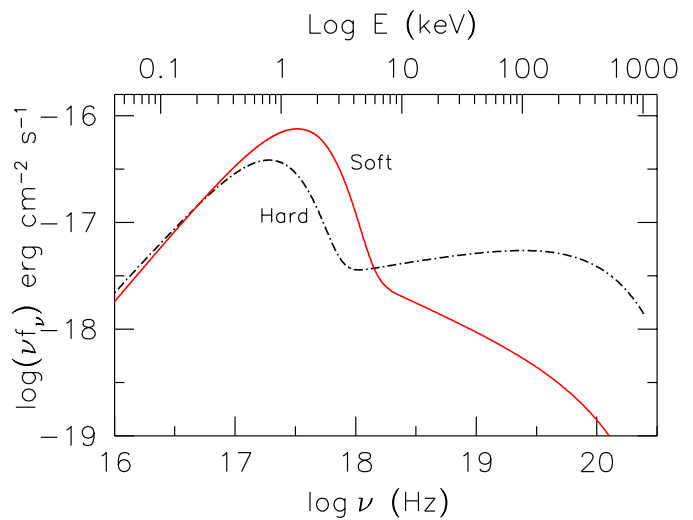}
\caption{The SEDs corresponding to the Soft and Hard states of the outburst of a
black hole of $10 M_{\odot}$. The two important components of the SED, namely,
the disk spectrum and the power-law have been added following the scheme
described in \citet{remillard06}. See \subsecn{subsec:SED} for the details.}
\labfig{fig:SED}
\end{center}
\end{figure}
%%%%%%%%%%%%%%%%%%%%%%%%%%%%%%%%%%%%%%%%%%%%%%%%%%%%%%%%%%%%%%%%%%%%%%%%%%%%%%

The SED of BHBs usually comprises of two components - (1) a thermal component
and (2) a non thermal power-law component with a photon spectrum $N(E) \propto
E^{-\Gamma}$ \citep{remillard06}. The thermal component is believed to be the
radiation from the inner accretion disk around the black hole, and is
conventionally modeled with a multi-temperature blackbody often showing a
characteristic temperature ($T_{in}$) near 1 keV.  During their outbursts the
BHBs transition through different states where the SED shows varying degrees of
contribution from the aforementioned components. The state where the radiation
from the inner accretion disk dominates and contributes more than 75\% of the
2-20 keV flux, is fiducially called the Soft state \citep{remillard06}. On
the other hand the fiducial Hard state is one when the non thermal power-law
contributes more than 80\% of the 2-20 keV flux \citep{remillard06}. For any
given BHB, the accretion disk usually appears to be fainter and cooler in this
Hard power-law state than it is in the Soft thermal state. \\

The radiation from a thin accretion disk may be modeled as the sum of local
blackbodies emitted at different radii and the temperature $T_{in}$ of the
innermost annulus (with radius $r_{in}$) of accreted matter is proportional to
$\left[\dot m_{obs} / (M_{BH} r_{in}^3) \right]^{1/4}$
%%%%%%%%%%%%%%%%%%%
%\begin{equation} 
%T(r_{in}) = 6.3 \times 10^5 \left( \dot m_{SED} \right)^{\frac{1}{4}} \left(\frac {M_{BH}} {10^8M_{\odot}} \right)^{-\frac{1}{4}} \left(\frac{r_{in}}{2 r_g}\right)^{-\frac{3}{4}}\rm{K} 
%\labequn{eqn:DbbTemp}
%\end{equation}
%%%%%%%%%%%%%%%%%%%
\citep{peterson97,frank02} where the observational accretion rate $\dot
m_{obs}$ is defined as 
\begin{equation}
\dot m_{obs} = L_{rad}/L_{Edd},
\labequn{eqn:mdot_obs}
\end{equation}
$L_{rad}$ being the luminosity in the energy range 0.2 to 20 keV and $L_{Edd}$
being the Eddington luminosity. A standard model for emission from a thin
accretion disk is available as disk blackbody \citep[hereafter
diskbb,][]{mitsuda84, makishima86} in
XSPEC\footnote{http://heasarc.gsfc.nasa.gov/docs/xanadu/xspec/}
\citep{arnaud96}. We use the diskbb in version 11.3 of XSPEC to generate the
disk spectrum $f_{disk}(\nu)$, where $T_{in}$ is used as an input. The other
required input for diskbb, the normalisation, is proportional to $r_{in}^2$.
%%%%%%%%%%%%%%%%%%%
%\begin{equation}
%A_{dbb} = \left\{ \frac {r_{in}/\rm{km}} {D/(10\,\,\rm{kpc})} \right\}^2 \cos\theta
%\labequn{DbbNorm}
%\end{equation}
%%%%%%%%%%%%%%%%%%%
%for an observer at a distance $D$ whose line-of-sight makes and angle $\theta$
%to the normal to the disk plane. 
To $f_{disk}(\nu)$ we add a hard power-law with a high energy cut-off, yielding
%%%%%%%%%%%%%%%%%%%
\begin{equation}
f(\nu) = f_{disk} (\nu) + [A_{pl} \nu^{-\alpha}] \exp{^{-\frac{\nu}{\nu_{max}}}}
\labequn{eqn:FullSed}
\end{equation}
%%%%%%%%%%%%%%%%%%%
to account for the full SED. We use the high energy exponential cut-off to
insert a break in the power-law at 100 keV.

We follow the prescription given in \citet{remillard06} to choose appropriate
values of the relevant parameters to derive the two representative SEDs for a
black hole of $10 M_{\odot}$.  
\begin{enumerate}
\item[$\bullet$] {\bf Soft state} (\fig{fig:SED} solid red curve): In the Soft
state the accretion disk extends all the way to $r_{in} = 3R_s = 6r_g$. Thus
$T_{in} = 0.56 \kev$. The power-law has $\Gamma = 2.5$ and $A_{pl}$ is chosen
in such a way that disk contributes to 80\% of the 2-20 keV flux. 
\item[$\bullet$] {\bf Hard state} (\fig{fig:SED} dotted-and-dashed black
curve): With $r_{in} = 6R_s = 12r_g$ we generate a cooler disk with $T_{in} =
0.33 \kev$. The power-law is dominant in this state with $\Gamma = 1.8$ so that
2-20 keV flux is only 20\%. 
\end{enumerate}

For a $10 M_{\odot}$ black hole, $L_{Edd} = 1.23 \times 10^{39} \rm{erg \,
s^{-1}}$. Using the aforementioned fiducial SEDs, we can derive $\dot m_{obs} =
0.14$ using the Soft SED and is equal to $0.07$ for the Hard SED. Thus for
simplicity we assume $\dot m_{obs} = 0.1$ for the rest of
this paper. 

It is important to note here, the distinction between the disk accretion rate
$\dot{m}$ (\dequn{eqn:n_scaling}{eqn:mdot}) mentioned above, and the observed
accretion rate $\dot m_{obs}$ which is more commonly used in the literature.
One can define, 
\begin{equation}
\dot m = \frac{2}{\eta_{acc}} \frac{\dot m_{obs}}{\eta_{rad}}
\labequn{eqn:mdot_n_mdotobs}
\end{equation}
where the factor $2$ is due to the assumption that we see only one of the two
surfaces of the disk. 

The accretion efficiency $\eta_{acc}\simeq r_g/2r_{in}$ depends mostly on the
black hole spin. For the sake of simplicity, we choose the Schwarzchild black
hole, so that $\eta_{acc} \sim 1/12$, both in Soft and Hard state.

The radiative efficiency, $\eta_{rad} = 1$ if the inner accretion flow is
radiatively efficient i.e. it radiates away all the power released due to
accretion. This is the case for a standard (i.e. geometrically thin, optically
thick) accretion disk and is satisfied in the Soft state when the standard
accretion disk extends all the way up to $r_{in} = 6 r_g$.  Thus $\dot m = 24
\dot m_{obs} = 2.4$. We acknowledge that $\eta_{rad}$ can be expected to be $<
1$ in the Hard state because the interior most parts of the accretion disk may
be more complex. In the Hard state, part of the accretion power could be
advected and not radiated (like in accretion dominated accretion flow, ADAF),
or ejected \citep[like in Jet Emitting Disks,][]{ferreira06}.  Instead of going
into detailed calculations of such kind of accretion disks, we accounted for
the resultant modifications in the Hard SED, by merely increasing the standard
accretion disk radius $r_{in}$ to $12 r_g$, keeping in mind that the inner part
of the flow could be filled by a different, radiatively less efficient,
accretion flow. But,the fact that $\eta_{rad} < 1$ (in the hard state), may be
balanced by the fact that $\dot m_{obs, \, \rm{Hard}}$ is slightly smaller
(0.07) than $\dot m_{obs, \, \rm{Soft}}$ (0.14), so that  $\dot m_{obs} /
\eta_{rad}$ remains the same for the Soft and the Hard states. Hence, for the
sake of simplicity, we assume that the same value of $\dot m = 2.4$ can be
retained for the Hard state.

%%%%%%%%%%%%%%%%%%%%%%%%%%%%%

%%%%%%%%%%%%%%%%%%%%%%%%%%%%%%%%%%%%%%
\subsection{Constraints from atomic physics}
\labsubsecn{subsec:AtomicPhysicsConstraints}

%%%%%%%%%%%%%%%%%%%%%%%%%%%%%%%%%%%%%%%%%%%%%%%%%%%%%%%%%%%%%%%%%%%%%%%%%%%%%
\begin{figure}
\begin{center}
\includegraphics[scale = 1, width = 8 cm, trim = 0 0 0 0, clip, angle = 0]{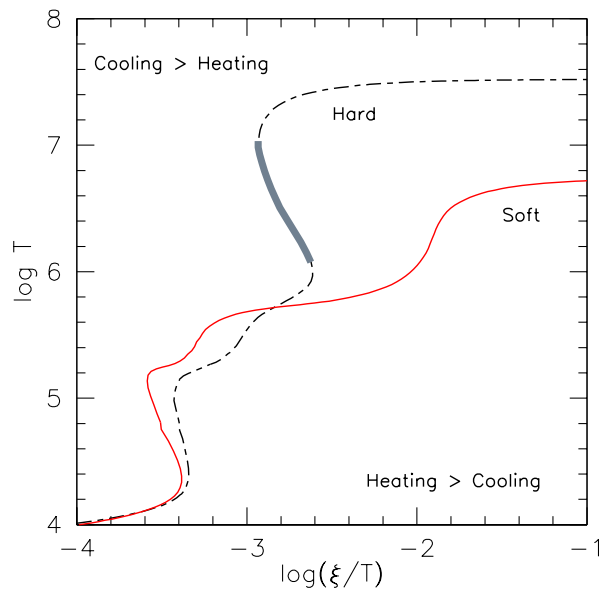}
\caption{The stability curves for photoionised gas with Solar abundance, $n_H =
10^{10} \,\, \rm{cm^{-3}}$ and $N_H = 10^{23} \,\, \rm{cm^{-2}}$ being
illuminated by the Soft and the Hard SEDs. A part of the Hard stability curve
is highlighted in thick gray - this is the negative slope part of the curve and
corresponds to unstable thermodynamic equilibrium. Gas with $\log \xi$ in this
part of the curve is unlikely to exist in nature. The Soft curve is stable in
the relevant part ($\log T \ge 5.5$) Note that both curves have a part with
negative slope at $\log T \lesssim 5.0$. However, this part of the stability
curve has such low values of $\log \xi \,\, (< 2.0)$ which are not relevant for
gas around BHBs. } 
\labfig{fig:Scurve}
\end{center}
\end{figure}
%%%%%%%%%%%%%%%%%%%%%%%%%%%%%%%%%%%%%%%%%%%%%%%%%%%%%%%%%%%%%%%%%%%%%%%%%%%%%%

%%%%%%%%%%%%%%%%%%%%%%%%%%%%%%%%%%%%%%%%%%%%%%%%%%%%%%%%%%%%%%%%%%%%%%%%%%%%%
\begin{figure}
\begin{center}
\includegraphics[scale = 1, width = 9 cm, trim = 0 240 0 0, clip, angle = 0]{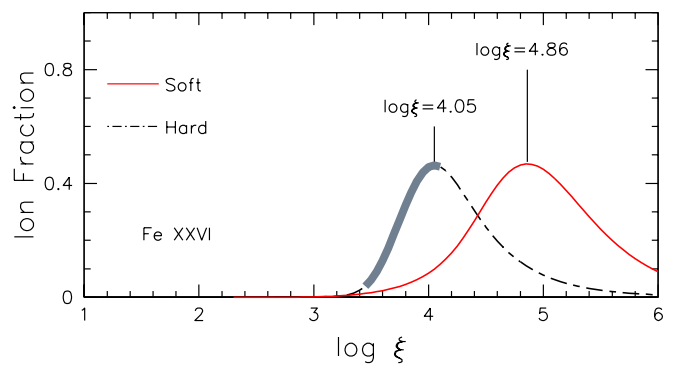}
\caption{The ion fraction distribution of FeXXVI with respect to $\log \xi$ is
shown for the two SEDs, Soft and Hard. The peak of the distribution is marked
and the corresponding $\log \xi$ values are labeled. Note that for the Hard
SED, a part of the distribution is highlighted by thick gray line - corresponds
to the thermodynamically unstable range of $\xi$. } 
\labfig{fig:If}
\end{center}
\end{figure}
%%%%%%%%%%%%%%%%%%%%%%%%%%%%%%%%%%%%%%%%%%%%%%%%%%%%%%%%%%%%%%%%%%%%%%%%%%%%%%

The MHD solutions can be used to predict the presence of outflowing material
over a wide range of distances. For any given solution, this outflowing material
spans large ranges in physical parameters like ionization parameter, density,
column density, velocity and timescales. Only part of this outflow will be
detectable through absorption lines - we refer to this part as the ``detectable
wind''.

%Ionization parameter 
Ionization parameter is one of the key physical parameters in determining which
region of the outflow can form a wind. There are several forms of ionization
parameter in the literature. In this paper we use the definition, more commonly
used by X-ray high resolution spectroscopists, namely $\xi = L_{ion}/(n_H
R^2_{sph})$ \citep{tarter69}, where $L_{ion}$ is the luminosity of the ionizing
light in the energy range 1 - 1000 Rydberg (1 Rydberg = 13.6 eV) and $n_H$ is
the density of the gas located at a distance of $R_{sph}$.  We assume that at
any given point within the flow, the gas is getting illuminated by light from a
central point source. This simplified approach is not a problem unless the wind
is located at distances very close to the black hole ($\leq 100 r_g$). The SEDs
for this radiation has been discussed in the previous \subsecn{subsec:SED}.  

For detecting the presence of ionized gas, we need to evaluate if the
ionization parameter of the gas is thermodynamically stable. Any stable
photoionised gas will lie on the thermal equilibrium curve or `stability' curve
of $\log T$ vs $\log(\xi/T)$ (\fig{fig:Scurve}). This curve is often used to
understand the structure of absorbing gas in AGN \citep[][and references
therein]{chakravorty08, chakravorty09, chakravorty12} and BHBs
\citep{chakravorty13, higginbottom15}. If the gas is located (in the $\xi - T$
space) on a part of the curve with negative slope then the system is considered
thermodynamically unstable because any perturbation (in temperature and
pressure) would lead to runaway heating or cooling. Gas lying on the part of
the curve with positive slope, on the other hand, is thermodynamically stable
to perturbations and hence likely to be detected when they will cause
absorption lines in the spectrum. 

With version C08.00 of CLOUDY\footnotemark \citep[][]{ferland98}.
\footnotetext{URL: http://www.nublado.org/ }
we generated stability curves using both the Soft and the Hard SEDs as the
ionizing continuum. For the simulation of these curves we assumed the gas to
have solar metallicity, $n_H = 10^{10} \,\, \rm{cm^{-3}}$ and $N_H = 10^{23}
\,\, \rm{cm^{-2}}$. Assuming these representative average values of $n_H$ and
$N_H$ are reasonable because the stability curves remain invariant when these
two parameters are varied over a wide range spanning several decades
\citep[see][for details]{chakravorty13}. The Soft stability curve (solid red
line in \fig{fig:Scurve}) has no unstable region, whereas the Hard one
(dotted-and-dashed black line) has a distinct region of thermodynamic
instability which is marked by the thick grey line.  This part of the curve
corresponds to $3.4 < \log \xi < 4.1$. Thus, this range of ionization parameter
has to be considered undetectable, when we are using the Hard SED as the source
of ionising light. 

Literature survey shows that it is usually absorption lines from H- and He-like
Fe ions that are detected (e.g \citealt{lee02}, \citealt{neilsen09},
\citealt{miller04}, \citealt{miller06}, \citealt{king12}). In fact, it is the
absorption line from FeXXVI that is most commonly cited as observed. A very
important compilation of detected winds in BHBs was presented in
\citep{ponti12}, and this paper also, concentrates the discussion around the
line from FeXXVI. Hence we choose the presence of the ion FeXXVI as a proxy for
detectable winds. The probability of presence of a ion is measured by its ion
fraction.  The ion fraction $I(X^{+i})$ of the $X^{+i}$ ion is the fraction of
the total number of atoms of the element $X$ which are in the $i^{\rm{th}}$
state of ionization. Thus, $$I(X^{+i}) = \frac{N(X^{+i})}{f(X) \,
N_{\rm{H}}},$$ where $N(X^{+i})$ is the column density of the $X^{+i}$ ion and
$f(X) = n(X)/n_{\rm{H}}$ is the ratio of the number density of the element $X$
to that of hydrogen. \fig{fig:If} shows the ion fraction of FeXXVI calculated
using CLOUDY. The ion fractions are of course, different based on whether the Soft
or the Hard SED has been used as the source of ionization for the absorbing
gas. The value of $\log \xi$, where the presence of FeXXVI is maximised,
changes from 4.05 for the Hard state, by $\sim$ 0.8 dex, to 4.86 for the Soft
state.

In the light of all the above mentioned observational constraints, we will
impose the following physical constraints on the MHD outflows (in
\dsecn{sec:EpPVar}{sec:WarmSolns}) to locate the detectable wind region within
them:  
\begin{itemize}
\item In order to be defined as an outflow, the material needs to have positive
velocity along the vertical axis ($z_{cyl}$).  
\item Over-ionized gas cannot cause any absorption and hence cannot be
detected. Thus to be observable via FeXXVI absorption lines we constrain the
material to have an upper limit for its ionization parameter. We imposed that
$\xi \leq 10^{4.86} \,\, {\rm{erg \, cm}}$ (peak of FeXXVI ion fraction) for
the Soft state. The ion fraction of FeXXVI peaks at $\xi = 10^{4.05}$ for
the Hard state, but this value is within the thermodynamically unstable range.
Hence for the Hard state, the constraint is $\xi \leq 10^{3.4} \,\, {\rm{erg \,
cm}}$, the value below which the thermal equilibrium curve is stable. 
\item The wind cannot be Compton thick and hence we impose that the integrated
column density along the line of sight satisfies $N_H < 10^{24}
{\rm{cm^{-2}}}$.
\end{itemize}

%%%%%%%%%%%%%%%%%%%%%%%%%%%%%%%%%%%%%%

%%%%%%%%%%%%%%%%%%%%%%%%%%%%%%%%%%%%%%
\subsection{Finding the detectable wind within the MHD outflow}
\labsubsecn{subsec:FindWind}

%%%%%%%%%%%%%%%%%%%%%%%%%%%%%%%%%%%%%%%%%%%%%%%%%%%%%%%%%%%%%%%%%%%%%%%%%%%%%
\begin{figure*}
\begin{center}
\includegraphics[scale = 1.0, height = 13.5 cm, trim = 0 150 0 250, angle = 90]{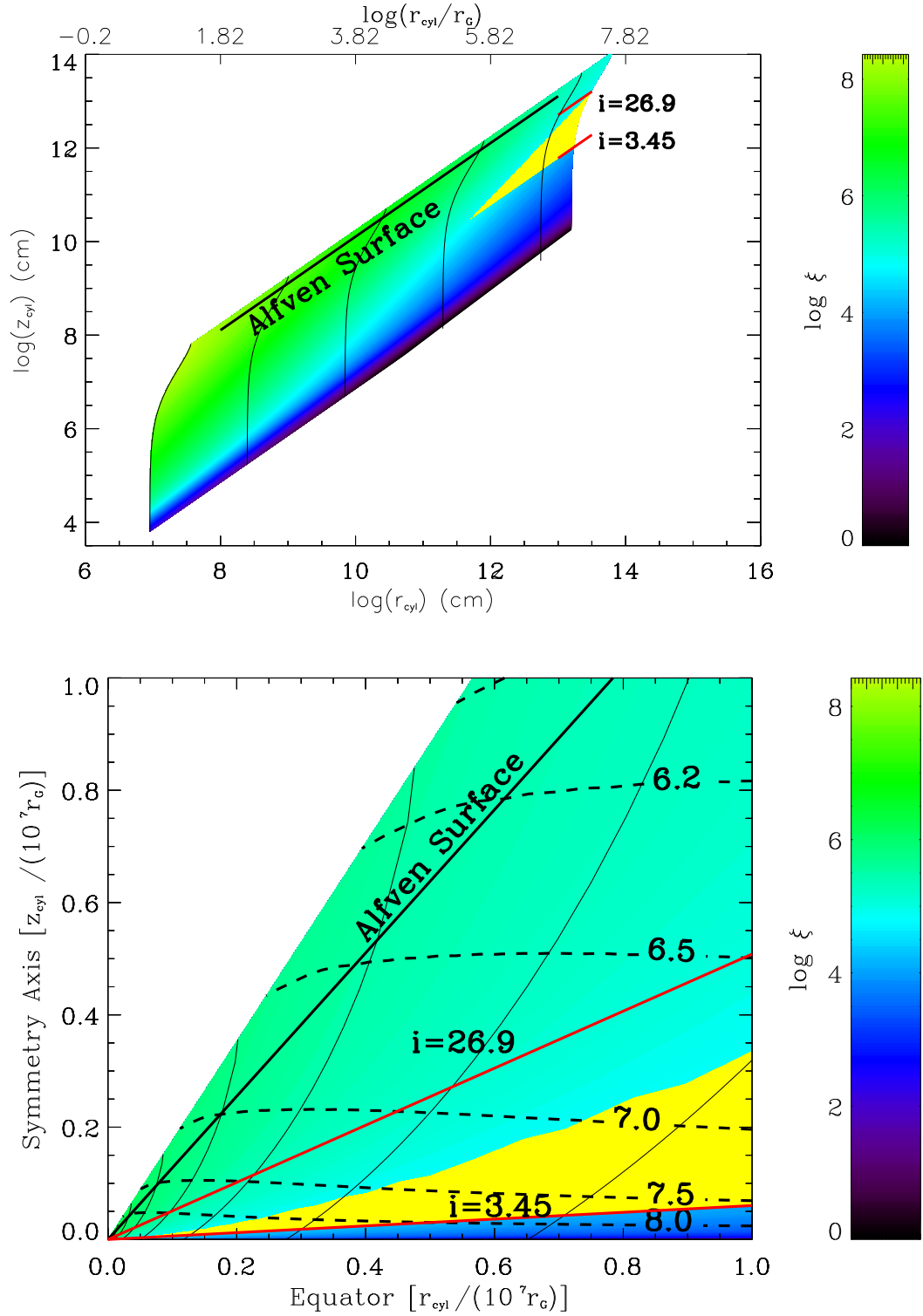}
\caption{\emph{Top Panel}: The distribution of the `Best Cold Set' in the
logarithmic plane of the radial ($r_{cyl}$) and vertical ($z_{cyl}$) distance
(in cylindrical co-ordinates) from the black hole. The distances are also
expressed in terms of the gravitational radius $r_g$ (top axis), which is $1.5
\times 10^6 \rm{cm}$ for a $10 M_{\odot}$ black hole. The colour gradient
informs about the $\xi$ distribution of the flow. The solid black lines
threading through the distribution show some of the magnetic field lines along
which material is outflowing. The Alfv\'en surface corresponding to the
solution is also marked and labelled. The yellow wedge highlights the wind part
of the flow - this material is optically thin with $N_H < 10^{24} \rm{cm^{-3}}$
and has sufficiently low ionization parameter (with $\xi < 10^{4.86} \rm{erg
\, cm}$) to cause FeXXVI absorption lines. The angular extent of the wind is
also clearly marked, where $i$ is the equatorial angle. \emph{Bottom Panel}: A
close up view of the wind region. The distances are expressed in linear scale,
but normalised to $10^7 r_g$. The dashed lines show the iso-contours of $n_H$,
while the associated labels denote the value of $\log n_H (\rm{cm^{-3}})$. }  
\labfig{fig:BestCold}
\end{center}
\end{figure*}
%%%%%%%%%%%%%%%%%%%%%%%%%%%%%%%%%%%%%%%%%%%%%%%%%%%%%%%%%%%%%%%%%%%%%%%%%%%%%

In this subsection we demonstrate how we choose the part of the MHD outflow
which will be detectable through absorption lines of FeXXVI.  For the
demonstration we use the MHD solution with $\varepsilon = 0.001$ and $p = 0.04$
which is illuminated by the Soft SED. Hereafter we will refer to this set of
parameters as the ``Best Cold Set''.  For the purpose of discussion in this
subsection, we will work with the Soft SED only, but in subsequent sections
additional calculations will be carried out for the scenario where the MHD
outflow is illuminated by the Hard SED.

The Best Cold MHD solution provides us with the knowledge of the density of the
material at each point within the outflow. Further, we know the Soft SED (both
shape and intensity). Hence at each point in the outflow we can calculate the
ionization parameter $\xi = L_{ion}/(n_H R^2_{sph})$. \fig{fig:BestCold} shows
the ionization parameter distribution (colour gradient) and the density
distribution (iso-density contours on the lower panel) of the outflow due to
the ``Best Cold Set''.  The solid black lines threading through the
distribution shows the magnetic field lines along which material is outflowing.
The MHD solutions are mathematically self-similar in nature, which essentially
means that we can propagate the solutions infinitely. However we have
restricted the last streamline to be anchored at $r_o = 10^7 r_g$. The top
panel of the figure is a global view, which shows the entire span of the MHD
solution that has been evaluated.

To find the wind region (detectable through FeXXVI absorption lines) within
this outflow we have to impose the three required physical conditions listed in
\subsecn{subsec:AtomicPhysicsConstraints}. The resultant wind region is
highlighted as the yellow wedge in \fig{fig:BestCold}. We see that the wind is detected
only from the outer parts of the flow with $\log R_{sph}|_{\rm{wind}} / r_g
\geq 5.4$. The lowest and highest equatorial angle ($i$) of the line of sight
are clearly marked for the wind region (in both panels). The observer will have
to view the BHB within this angular range to be able to detect the wind. The
wind is equatorial, for the `Best Cold Set', not extending beyond $i =
26.9^{\circ}$.  In the lower panel of \fig{fig:BestCold} we use a linear (but
normalised by $10^7 r_g$) scale for $r_{cyl}$ and $z_{cyl}$, which renders us a
close up view of the wind region within the solution. The labelled dashed black
lines are the iso-contours for the number density $\log n_H (\rm{cm^{-3}})$.
The velocities $v_{obs}$ (not shown in the Figure) in this region fall within
the range $10^2 - 10^3 \rm{km\, s^{-1}}$.

%%%%%%%%%%%%%%%%%%%%%%%%%%%%%%%%%%%%%%%%%%%%%%%%%%%%%%%%%%%%%%%%%%%%%%%%%%%%%
\begin{figure*}
\begin{center}
\includegraphics[scale = 1, height = 19 cm, trim = 70 10 125 20, clip, angle = 90]{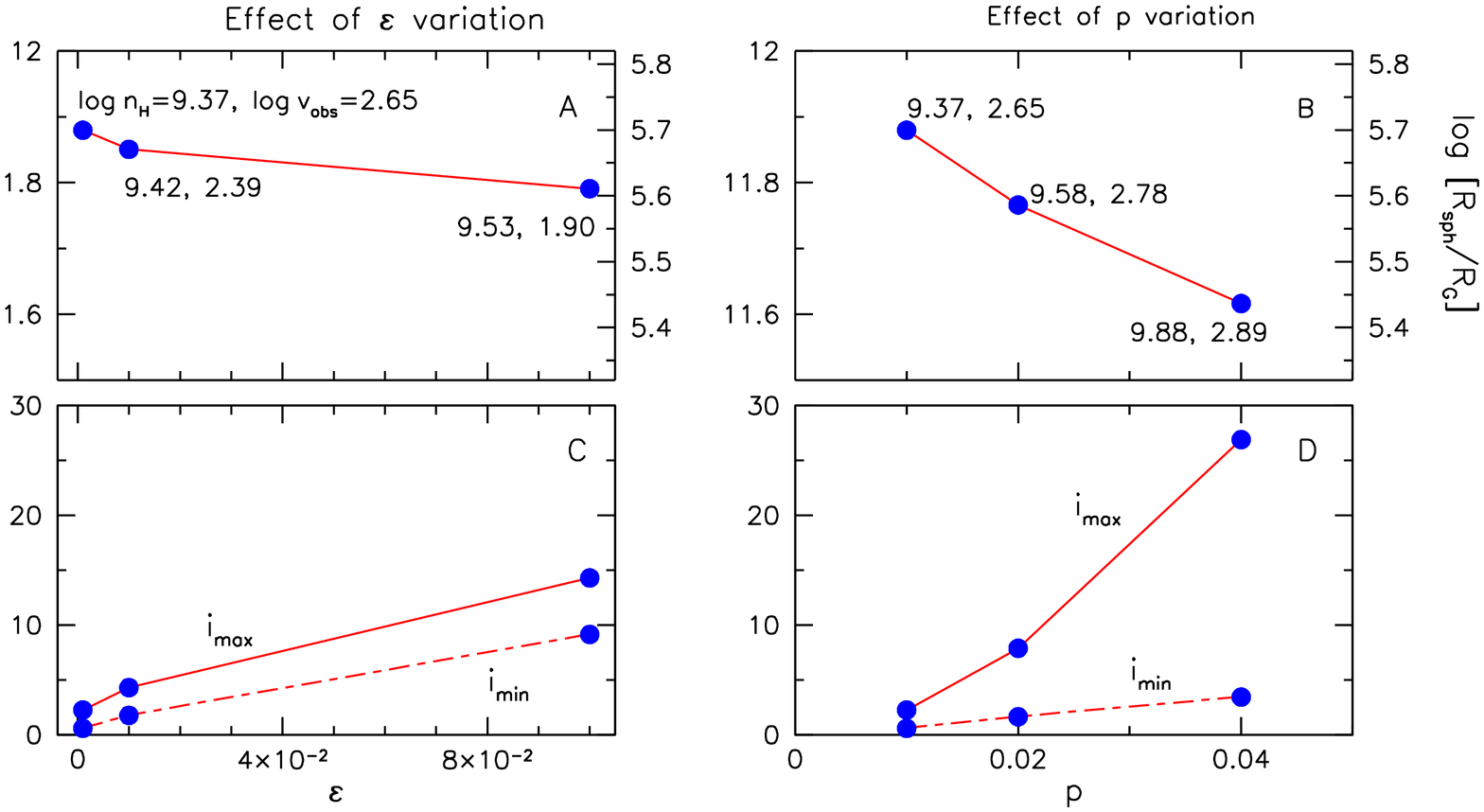}
\caption{The physical parameters of the wind are plotted as a function of
$\varepsilon$ (left panels) and $p$ (right panels), while using the Soft SED as
the ionizing continua. {\emph Top Panels}: For the closest wind point, we plot
the logarithm of $R_{sph}|_{wind}$ in the left panel A as a function of the
disk aspect ratio $\varepsilon$ and as a function of the accretion index $p$ in
the right panel B. $p = 0.01$ is held constant for the solutions in the left
panels and $\varepsilon = 0.001$ is kept constant for those in the right
panels. Each blue circle in the figure represents a MHD solution. The logarithm
of two other relevant quantities, $n_H$ and $v_{obs}$ for the closest wind
point are labeled at each point - these are their maximum possible values
within the wind region, for a given MHD solution. {\emph Bottom Panels}: The
minimum ($i_{min}$) and the maximum ($i_{max}$) equatorial angles of the line
of sight, within which the wind can be observed, is plotted as a function of
$\varepsilon$ (left) and of $p$ (right).} 
\labfig{fig:RsphDeltai}
\end{center}
\end{figure*}
%%%%%%%%%%%%%%%%%%%%%%%%%%%%%%%%%%%%%%%%%%%%%%%%%%%%%%%%%%%%%%%%%%%%%%%%%%%%%%

This same method of finding the wind, and the associated physical conditions is
used for all the cold MHD solutions considered in this paper. In the subsequent
sections we will vary the MHD solutions (i.e. $\varepsilon$ and $p$) and
investigate the results using both the Soft and Hard SEDs. 

To ensure that the wind is in thermal equilibrium, it is important to compare
the various physical timescales. We used CLOUDY to evaluate the cooling time
scales at each point within the wind region of the solution.  CLOUDY assumes
that atomic processes (including photoionization and recombination cooling)
occur on timescales that are much faster than other changes in the system, so
that atomic rates have had time to become ``time-steady''. These atomic
processes, in addition to some other continuum processes like Comptonization
and Bremsstrahlung, are responsible for heating and cooling the gas. Whether
the atomic processes dominate over the continuum processes is determined by the
ionization state and/or the temperature of the gas. For photoionized wind we
expect the atomic processes to dominate.  However, one way to make sure that
the gas satisfies the time-steady condition (which is assumed by CLOUDY) is to
check the CLOUDY computed cooling time scale against the dynamical time scales
from our physical MHD models. CLOUDY defines the cooling time scale as the time
needed to loose half of the heat generated in the gas due to various atomic and
continuum processes. Thus thermal equilibrium is also ensured as long as the
cooling time scale is smaller than the dynamical time scale $\tau_{dyn}$ -
which was found to be true within the wind region of the outflow for the `Best
Cold Set'.

%%%%%%%%%%%%%%%%%%%%%%%%%%%%%%%%%%%%%%

%%%%%%%%%%%%%%%%%%%%%%%%%%%%%%%%%%%%%%%%%%%%%%%%%%%%%%%%%%%%%%%%%%%%%%%%%%%%%%
%\section{A phenomenological probe of the ``cold'' MHD solutions}
\section{The cold MHD solutions}
\labsecn{sec:EpPVar}

%%%%%%%%%%%%%%%%%%%%%%%%%%%%%%%%%%%%%%
\subsection{Effect of variation of the parameters of the MHD flow}
\labsubsecn{ParameterVar}

Here we aim to find which of the two parameters $\varepsilon$ and $p$ is more
influential in producing the wind. The value of $p$ and $\varepsilon$ decides
the density of material at the launching point of our magneto-hydrodynamic
outflow (\equn{eqn:rho+}). The extent of magnetisation in the outflow is also
dependant on $p$ (\secn{sec:MhdWinds}). It is these two parameters that links
the density and other physical properties of the outflow with the accretion
disk. Since a particular pair of $p$ and $\varepsilon$ will result in a unique
MHD solution, we can generate different MHD solutions, by changing the values
of $p$ and $\varepsilon$. On each of these solutions, we perform the methods
described in the previous \subsecn{subsec:FindWind} and investigate the wind
part of the outflow. 

To judge the influence of $p$ and $\varepsilon$, in a quantitative way, we
compare some physically relevant parameters of the wind. For observers, one
important set of wind parameters are the distance, density and velocity of the
point of the wind closest to the black hole. Hereafter we shall call this point
as the `closest wind point'.  Another quantity of interest would be the
predicted minimum and maximum angles (of the line of sight) within which the
wind can be observed. We conduct this exercise using both the SEDs - Soft and
Hard. The results are plotted in \fig{fig:RsphDeltai}.
 
The exact value of these quantities should not be considered very rigorously,
because the value is decided by the various constraints that we have applied.
It is more important to note the changes in these quantities as $\varepsilon$
and $p$ vary. The relative changes should be used to assess how variations in
$\varepsilon$ and $p$ increase the possibilities of detecting the wind.

%%%%%%%%%%%%%%%%%%%
\subsubsection{Variation of the disk aspect ratio $\varepsilon$}
\labsubsubsecn{subsubsec:EpVar}

For the closest wind point, we plot $R_{sph}|_{wind}$ versus the value of
$\varepsilon$ of the MHD solution, in panel A of \fig{fig:RsphDeltai}. Further,
$n_H$ and $v_{obs}$ for this point are labelled. {\it Using the Soft SED}, the
closest wind point reaches closer to the black hole by a factor of 1.06 as
$\varepsilon$ increases from 0.001 to 0.01, and then by a farther factor of
1.14 as $\varepsilon$ increases to 0.1. The density at the closest point is
$n_H|_{max} = 10^{9.37} \rm{cm^{-3}}$, for $\varepsilon = 0.001$ . Note that
for any given solution, the density at the closest point is the maximum
attainable density within the wind region, for that particular MHD solution.
This maximum attainable density of the wind, increases as $\varepsilon$
increases to 0.01 and then to 0.1. However, as a function of $\varepsilon$, the
variation in this quantity is not very high, but only 0.16 dex. Like density,
for a given solution, the velocity at the closest wind point, $v_{obs}|_{max}$,
is the highest that can be attained by the detectable wind. This quantity
monotonically decreases by 0.26 dex and then by 0.49 dex as $\varepsilon$
increases from 0.001 to 0.01 and then to 0.1. This means, to get winds with
higher speed, we need disks with higher aspect ratios.

$i_{min}$ is the minimum and $i_{max}$ the maximum equatorial angles of the
line of sight, within which the wind can be detected. Panel C of
\fig{fig:RsphDeltai} shows the changes in the angles as $\varepsilon$ varies.
One can easily judge the angular extent of the wind by gauging the difference
between $i_{min}$ and the maximum angles, for a particular solution (and SED).
$i_{min}$ rises from 0.60 to 1.78 to 9.15 and $i_{max|{Soft}}$ increases from
2.27 to 4.31 to 14.3 as $\varepsilon$ varies from 0.001 to 0.01 to 0.1. The
growth of $\Delta i = i_{max} - i_{min}$ with $\varepsilon$ shows that the wind
gets broader as the disk aspect ratio increases. 

%%%%%%%%%%%%%%%%%%%

%%%%%%%%%%%%%%%%%%%
\subsubsection{Variation of the ejection index $p$}
\labsubsubsecn{subsubsec:pVar}

As $p$ increases, the wind moves closer to the black hole (panel B of
\fig{fig:RsphDeltai}) - $R_{sph}|_{wind}$ drops by a factor of 1.3 as $p$ goes
from 0.01 to 0.02 and then reduces further by a factor of 1.41 when $p$ is
increased to 0.04, while using the Soft SED. The total change in the density of
the closest wind point is 0.51 dex as $p$ changes from 0.01 to 0.04. Thus both
$R_{sph}|_{wind}$ and $n_H|_{max}$ are effected more by the variation in $p$
than by the variation in $\varepsilon$ (within the range of these parameters
investigated by us). The velocity $v_{obs}|_{max}$ of the closest point
however, varies far less with change in $p$, the total decrease being only 0.24
dex.

$i_{min}$ and $i_{max}$ as a function of $p$ (using the Soft SED) is shown in
panel D of \fig{fig:RsphDeltai}.  As $p$ goes from 0.01 through 0.02 to 0.04,
the minimum angle rises from 0.60 through 1.65 to 3.45, a range rather smaller
than that caused by the $\varepsilon$ variation. $i_{max}$ goes from 2.27 to
7.89 to 26.9. Thus the growth of $\Delta i$ is rendered to be higher as a
function of increase in $p$, implying a higher probability of detecting a wind
when the flow corresponds to higher $p$ values.  Since $p$ is the relatively
more dominant (compared to $\varepsilon$) disk parameter in increasing the
density at a given distance, the resultant outflowing material has lower
ionisation. This is a favourable influence to cause detectable winds.

%%%%%%%%%%%%%%%%%%%

%%%%%%%%%%%%%%%%%%%%%%%%%%%%%%%%%%%%%%
\subsection{Cold solutions for the Hard state} 
\labsubsubsecn{subsubsec:ColdHard}

For the entire range of $\varepsilon$ (0.001 - 0.1) and $p$ (0.01 - 0.04) we
analysed the MHD solutions illuminated by the Hard SED, as well. Note that for
the Hard SED, we have to modify the upper limit of $\xi$ according to the
atomic physics and thermodynamic instability considerations
(\subsecn{subsec:AtomicPhysicsConstraints}). With the appropriate condition,
$\log \xi \le 3.4$, we could not find any wind portions within the Compton thin
part of the outflow, for any of the MHD solutions. 

This is a very significant result, because this provides strong support to the
observations that BHBs do not have winds in the Hard state. We will discuss
this issue further complimented with better quantitative details in
\subsecn{subsec:WarmHard}.  

%%%%%%%%%%%%%%%%%%%%%%%%%%%%%%%%%%%%%%

%%%%%%%%%%%%%%%%%%%%%%%%%%%%%%%%%%%%%%

\subsection{The need for Warm MHD solutions} 
\labsubsubsecn{subsubsec:NeedWarmSoln}

The density reported for most of the observed BHB winds $\ge 10^{11}
\,\rm{cm^{-3}}$ and the distance estimates place the winds at $\le 10^{10} \,
\rm{cm}$ \citep{schulz02, ueda04, kubota07, miller08, kallman09}. Our analysis
in the previous subsections show that $R_{sph}|_{wind}$ is too high and
$n_H|_{max}$ is too low even for the `Best Cold Solution' to match
observations. The purpose of this section is to understand which parameter of
the accretion-ejection process can provide us with a MHD solution capable of
explaining observed (or derived) parameters of BHB winds. Studying the effect
of the disk parameters gives us a clear indication that increasing the value of
the ejection index $p$ favours the probability of detecting winds, as
demonstrated by the larger extent of increase in $\Delta i$. Further, the
increase in $p$ results in two more favourable effects - the closest wind point
moves closer to the black hole and causes a higher increase in density.

The above phenomenological tests of the $\varepsilon - p$ space, indicates that
a MHD solutions with higher $\varepsilon$, say 0.01, and a high $p \ge 0.04$
would be the better suited to produce detectable winds, comparable to
observations.  However there are limitations on the $\varepsilon - p$
combination imposed by the physics of the MHD solutions (see
\fig{fig:paramF97}) and it is not possible to reach larger values of $p$ for
the cold solutions with isothermal magnetic surfaces.  As shown in
\citet{casse00b}, to get denser outflows with larger $p$, some additional
heating needs to take place at the disk upper layers leading to a warming up of
the wind. The authors argued that the origin of this extra heating could be due
to illumination from an external source or enhanced turbulent dissipation
within the disk surface layers.  Let us now investigate in the following
section if a warm solution is indeed, much better for producing winds matching
observations.

%%%%%%%%%%%%%%%%%%%

%%%%%%%%%%%%%%%%%%%%%%%%%%%%%%%%%%%%%%

%%%%%%%%%%%%%%%%%%%%%%%%%%%%%%%%%%%%%%%%%%%%%%%%%%%%%%%%%%%%%%%%%%%%%%%%%%%%%

%%%%%%%%%%%%%%%%%%%%%%%%%%%%%%%%%%%%%%%%%%%%%%%%%%%%%%%%%%%%%%%%%%%%%%%%%%%%%%
\section{Warm MHD solutions}
\labsecn{sec:WarmSolns}

%%%%%%%%%%%%%%%%%%%%%%%%%%%%%%%%%%%%%%%%%%%%%%%%%%%%%%%%%%%%%%%%%%%%%%%%%%%%%
\begin{figure*}
\begin{center}
\includegraphics[scale = 1.0, width = 16 cm, trim = 0 0 0 0, angle = 0]{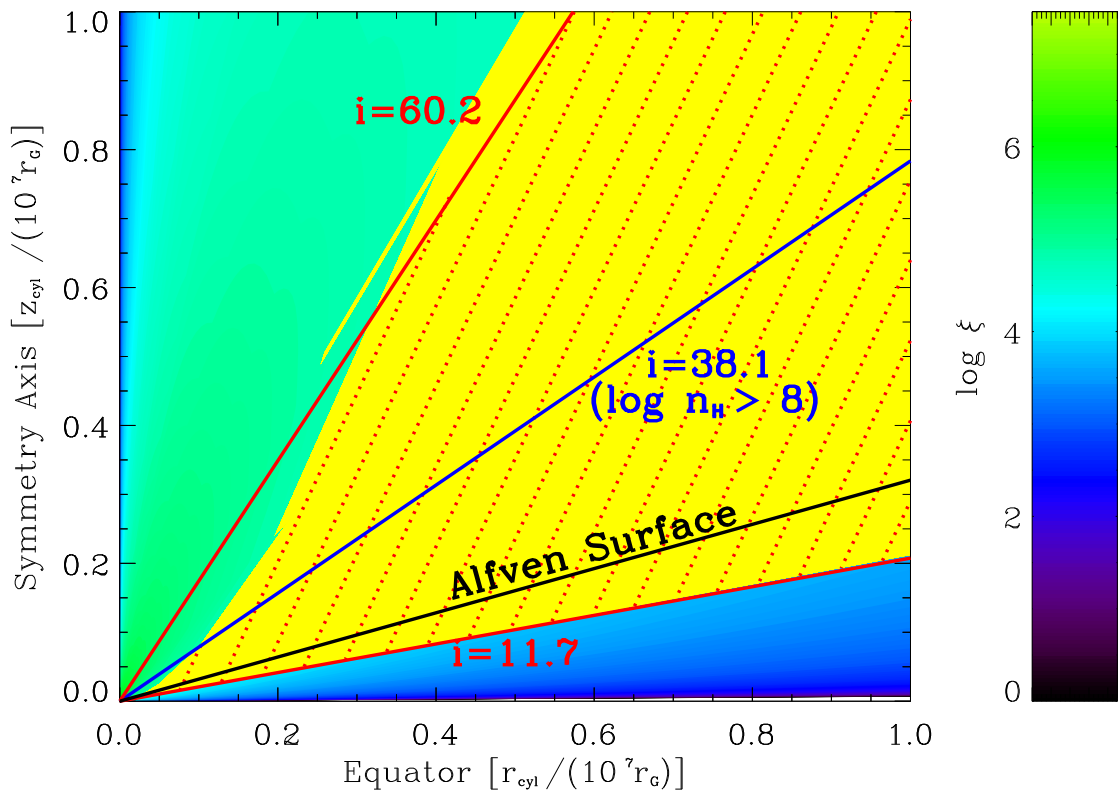}
\caption{The ionization parameter distribution for a Warm MHD solution with
$\varepsilon = 0.01$ and $p = 0.10$. The yellow region within the outflow is
obtained in the same way as in \fig{fig:BestCold}. The shaded region (with
dotted red lines) is the wind region within such a warm outflow - to obtain
this region we used the additional constraint that the cooling timescale of the
gas has to be lower than the dynamical time scale. Further, the solid blue line
with $i = 38.1^{\circ}$ is drawn to depict that high density material ($\log
n_H \ge 8.0$) in the flow is confined to low equatorial angles. }  
\labfig{fig:Warm}
\end{center}
\end{figure*}
%%%%%%%%%%%%%%%%%%%%%%%%%%%%%%%%%%%%%%%%%%%%%%%%%%%%%%%%%%%%%%%%%%%%%%%%%%%%%

In this section we investigate the properties of the wind as a function of
increasing $p$, but for warm MHD solutions. Here we choose to ignore the effect
of $\varepsilon$, because in the previous section we found that variation in
$\varepsilon$ (over two orders of magnitude) has very little effect on changing
the physical quantities of the wind.  Further, in the previous sections we
found that the wind does not exist for the Hard SED. Hence, we shall conduct
the extensive calculations with the Soft SED only. We shall however, discuss
winds in the Hard state in \subsecn{subsec:WarmHard} as a part of general discussions. 

%{\color{blue} {\Huge Done till here.}} 

Self-confined outflows require $\sigma^+ \simeq 1/p$ larger than unity, as
pointed out through  \equn{eqn:mag_sigma} in \subsecn{subsec:params}. Moreover,
the power in the outflow is always a sizable fraction of the mechanical power 
\begin{equation}
L_{acc} = \left [ \frac{G M_{BH} \dot M_{acc}(r)}{2 r} \right ]^{r_{in}}_{r_{out}}
\end{equation}
released by the accreting material between the inner radius $r_{in}$ and the
outer radius $r_{out}$. Because $\dot M_{acc}(r) \propto r^p$ in a disk,
launching outflows, one gets $L_{acc} =0$ for $p=1$.  This is why, unless there
is an external source of energy, $p=1$ is a maximum limit, and in fact,
powerful magnetically driven flows require a much smaller ejection index.  To
consider what highest value of $p$ should be aimed for, we scout the
literature. We find two relevant references, namely, \\ 
(a) \citet{casse00b} who computed warm MHD accretion-ejection solutions up to
$p=0.456$ to model winds mostly, in young stellar objects and \\
(b) a series of papers by Fukumura \etal \citep{fukumura10a, fukumura10b,
fukumura14, fukumura15}, who used a model with $p=0.5$.\\ 
Hence while attempting to generate the disk surface heated, magnetically driven
and magnetically confined outflows, we will limit ourselves to $p \le 0.5$.

For this paper, we obtain dense warm solutions (with higher values of $p$, i.e.
$p \ge 0.04$) through the use of an ad-hoc heating function acting along the
flow.  This additional heating needs to start within the disk itself, in the
resistive MHD layers, in order to cause a larger mass loading at the base of
the outflow. However, the heating requires to be maintained for some distance
within the outflow too, into the ideal MHD zone. This is necessary in order to
help the launching of these dense outflows and tap the thermal energy content
instead of the magnetic one \citep[refer to][for more details]{casse00b}. It
must therefore be realized that any given ``warm solution'' from a near
Keplerian accretion disk is based on an ad-hoc heating term \citep[the function
$Q$ in][, see also \secn{sec:ColdWarmSolns} of Appendix]{casse00b}. The
physical mechanism behind the heating term has not been specified. However, a
posteriori calcualtions show that even a few percents of the released accretion
energy would be enough to give rise to such warm MHD winds. Whether or not MHD
turbulence in accretion disks can provide such a surface heating is an open
theoretical issue. For further discussion on this see \secn{sec:ColdWarmSolns}
of the Appendix. On the other hand the heating could be caused by the
illumination form the interior parts of the disk. To determine and/or
distinguish between the physical cause of the heating is a rigorous theoretical
study in itself and is beyond the scope of this paper.
 
To ease comparison between various warm models, we use the same shape for the
heating function, while playing only with its normalization to increase $p$
\citep[the larger the heat input, the larger the value of $p$, see Figure 2
in][]{casse00b}. For $\varepsilon = 0.01$ we could achieve a maximum value of
$p=0.11$. For the purpose of this
paper, it is not required to provide the ``most massive'' (i.e.  largest
possible $p=0.5$) solution - it is enough to show general trends. However, we
are developing the methods to generate denser MHD solutions with $p = 0.5$ and
these solution(s) will be reported in our subsequent publications where we will
attempt to model the winds observed in specific outbursts of specific BHBs.  

%%%%%%%%%%%%%%%%%%%%%%%%%%%%%%%%%%%%%%%%%%%%%%%%%%%%%%%%%%%%%%%%%%%%%%%%%%%%%
\begin{figure*}
\begin{center}
\includegraphics[scale = 1.0, width = 16 cm, trim = 0 240 0 0, angle = 0]{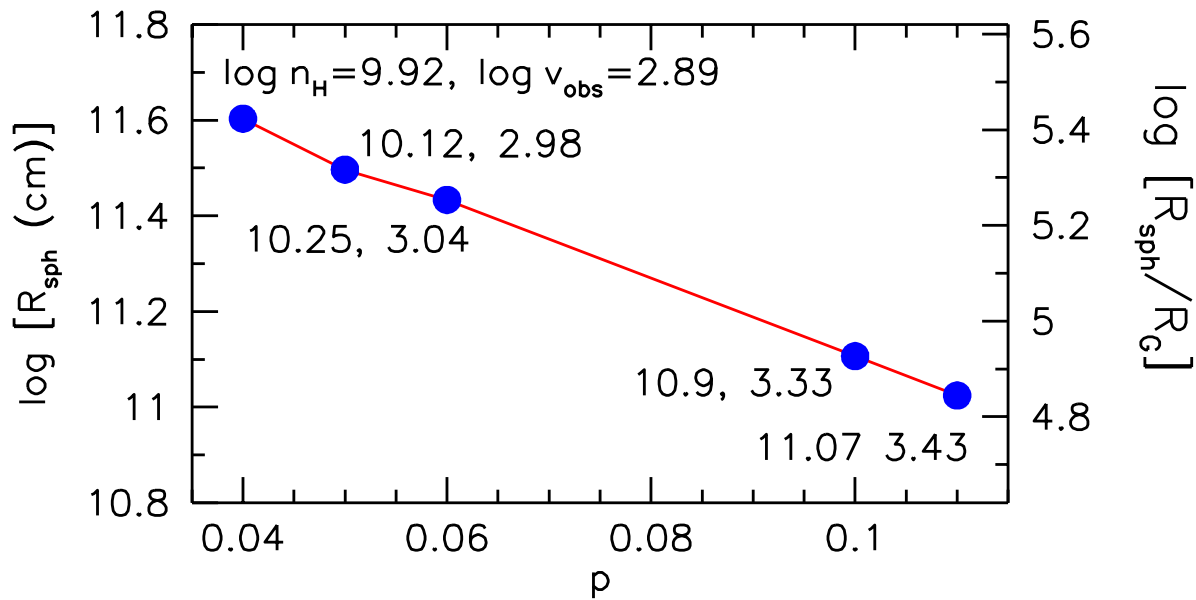}
\caption{Distance (density and velocity) of the closest wind point is (are)
plotted (labelled) as a function of $p$ for all the warm MHD solutions that we
investigated. $\varepsilon = 0.01$ is constant.}  
\labfig{fig:WarmPhen}
\end{center}
\end{figure*}
%%%%%%%%%%%%%%%%%%%%%%%%%%%%%%%%%%%%%%%%%%%%%%%%%%%%%%%%%%%%%%%%%%%%%%%%%%%%%

\fig{fig:Warm} shows the wind for a Warm MHD solution with a rather high $p =
0.10$. The wind (yellow region) spans a much wider range (than even the ``Best
Cold Solution'') and extends far beyond the Alfven surface which was not the
case for the cold MHD solutions. Hence we introduce an additional constraint
derived from timescale considerations.  The lower angular limit ($i =
11.7^{\circ}$) is derived due to the constraints of $\xi$ and $N_H$. Next, we
used CLOUDY to calculate the cooling timescales of the gas at each point within
the yellow wind region of the outflow. Note that for the timescale calculations
using CLOUDY (which are computationally expensive) we have used a much coarser
grid of $i$ than that used for other calculations of the MHD solutions. This is
sufficient for our purpose here, where a coarse upper limit on $i$ is
sufficient. To be consistent with a photoionised wind which is in thermal
equilibrium, the cooling timescale needs to be shorter than the dynamical
timescale. This timescale condition was satisfied within the yellow region if
$i \le 60^{\circ}$. Thus the red-dotted shaded region is the resultant
detectable wind. However, note that the densest parts of the wind is confined to
low equatorial angles. For example, gas with $n_H \ge 10^8 \, \rm{cm^{-3}}$
will lie below $i = 38.1^{\circ}$. For this solution, we further calculated
$B_z \sim 70$ Gauss (\equn{eqn:B_scaling}) at the disk mid plane at a distance
$r_{cyl} = R_{sph} = 1.28 \times 10^{11} \rm{cm}$.  

We investigated warm MHD solutions with a range of values of $p$. In
\fig{fig:WarmPhen} we have plotted the distance of the closest wind point for
all those solutions. Each point is also labelled with the respective values of
density and velocity. Between the $p = 0.04$ solution and the one with the
highest $p = 0.11$ (that we could achieve) $R_{sph}|_{wind}$ goes closer by a
factor of 3.79 and stands at $7.05 \times 10^4 \, r_g$. The highest density
that we could achieve is $\log n_H = 11.1$ and the highest velocity is $\log
v_{obs} = 3.43$. Hereafter we shall refer to the $\varepsilon = 0.01$ and $p =
0.10$ warm MHD solution as the ``Best Warm Solution''.

Clearly, warm solutions do a much better job than cold ones, as expected.
However, some observational results require the winds to have higher density
and lower distance than those produced by the ``Best Warm Solution''. In the
following section we discuss the possibilities in which we can theoretically
achieve more stringent values demanded by observations of some extreme winds.

%%%%%%%%%%%%%%%%%%%%%%%%%%%%%%%%%%%%%%%%%%%%%%%%%%%%%%%%%%%%%%%%%%%%%%%%%%%%%%

%%%%%%%%%%%%%%%%%%%%%%%%%%%%%%%%%%%%%%%%%%%%%%%%%%%%%%%%%%%%%%%%%%%%%%%%%%%%%%
\section{Discussions}
\labsecn{sec:discussion}

\subsection{Towards the extreme MHD winds}

%%%%%%%%%%%%%%%%%%%%%%%%%%%%%%%%%%%%%%
\subsubsection{Choice of upper limit of $\xi$}
\labsubsubsecn{subsubsec:UpperXi}

The ionization parameter is the key parameter in defining the wind region
within the outflow. Here we discuss (a) the possibility of changing $\xi$ if
$\dot m$ changes and (b) the effect if the upper limit of $\xi$ is changed. 

(a) In the definition of $\xi$, the density $n_H$ in the denominator is
proportional to $\dot m$ (see \equn{eqn:n_scaling}). In the numerator,
$L_{ion} \propto L_{rad}$ and we also assume $L_{rad}$ to be proportional to
$\dot m$ (see \subsecn{subsec:SED}). Hence for a given MHD solution, changing
$\dot m$ will not change the $\xi$ distribution within the outflow. 

In the case of inefficient accretion flow like ADAF, $L_{rad} \propto
\dot{m}^2$, and changes in $\dot m$ could have some effects. However, we are
considering here, physical scenarios, where the accretion disk is radiatively
efficient with $\dot{m}_{obs} \sim 0.1$. Hence, accepting $L_{rad} \propto \dot m$ is a reasonable assumption.

(b) We used the limit $\log \xi \le 4.86$ to define the detectable wind. Note that
for the Soft SED, $\log \xi = 4.86$ corresponds to the peak of the ion fraction
of FeXXVI (\fig{fig:If}). The ion can have significant presence at higher
$\xi$. For example, at $\log \xi = 6.0$ FeXXVI is still present, but at $\sim 1/4$ of
its peak value. Further, there are other ions (including NiXXVIII) which peak
at higher values of $\xi$ \citep[see Figure 4 of][]{chakravorty13}. Such ions
have been reported in \citet{miller08}. In fact such ions may be routinely
detected in data from the future X-ray telescopes like Astro-H and Athena. It
is thus instructive to investigate how the properties of the closest wind point
(for a given solution) are modified when the constraint on upper limit of $\log
\xi$ are changed.

For the best warm solution we calculated the physical parameters for the
closest wind point with a modified upper limit $\log \xi = 6.0$. We find that
$R_{sph}|_{wind}$ decreases by a factor of 93.4 bringing this point to $9.1
\times 10^2 r_g$. The density at this point is $\log n_H = 13.71$ and the
velocity is $\log v_{obs} = 4.28$. Thus we see that the parameters of closest
point is sensitively dependant on the choice of the upper limit of $\xi$. 

%%%%%%%%%%%%%%%%%%%%%%%%%%%%%%%%%%%%%%

%%%%%%%%%%%%%%%%%%%%%%%%%%%%%%%%%%%%%%
\subsubsection{The need for denser warm solution}
\labsubsubsecn{subsubsec:DenserWarmSoln}

From the analysis presented in \subsecn{ParameterVar} and \secn{sec:WarmSolns}
it is clear that MHD solutions with larger $p$ favour winds which are closer to
the black hole. Even for the densest solution discussed in this paper, with
$\varepsilon = 0.01$ and $p = 0.11$, we cannot predict a wind closer than $7.05
\times 10^4 \, r_g$ (for $\log \xi \le 4.86$) and denser than $\log n_H >
11.07$ . However \citet{miller08} discussed that the wind in GRO J1655-40 was
very dense, where $\log n_H \ge 12$, and hence had to be very close to the
black hole at $\sim 10^3 r_g$. Thus, to explain such extreme winds, we need
denser warm MHD solutions with higher $p$. 

%%%%%%%%%%%%%%%%%%%%%%%%%%%%%%%%%%%%%%%%%%%%%%%%%%%%%%%%%%%%%%%%%%%%%%%%%%%%%
\begin{figure}
\begin{center}
\includegraphics[scale = 1, width = 9 cm, trim = 0 125 0 75, clip, angle = 0]{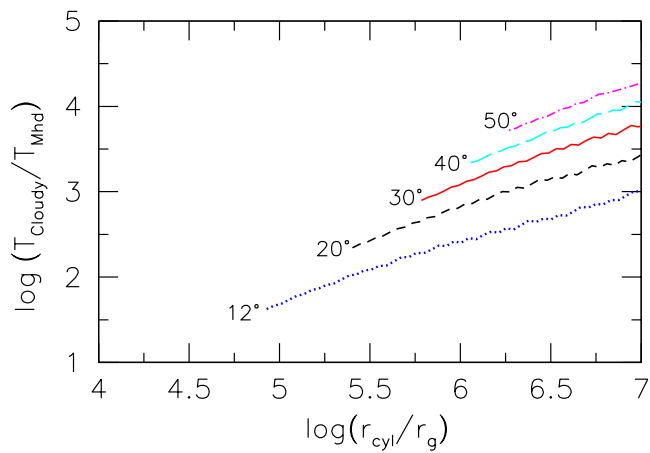}
\caption{Comparison of the different temperatures associated with wind in the
Best Cold Solution. $T_{Cloudy}$ is the temperature calculated using CLOUDY and
is the actual temperature of the photoionized gas. $T_{MHD}$ is the temperature
du to the MHD solution - the value that has been used to calculate all the
physical properties of the outflow. The ratio between these two temperatures
are plotted as a function of the distance from the black hole and at different
angles of line of sight. } 
\labfig{fig:TempCompare}
\end{center}
\end{figure}
%%%%%%%%%%%%%%%%%%%%%%%%%%%%%%%%%%%%%%%%%%%%%%%%%%%%%%%%%%%%%%%%%%%%%%%%%%%%%%

In the context of AGN, \citet{fukumura10a, fukumura10b, fukumura14, fukumura15}
have been able to reproduce the various components of the absorbing gas using
MHD outflows which would correspond to $p \simeq 0.5$. As discussed in
\secn{sec:WarmSolns}, we have not been able to reproduce such high values of
$p$ and are limited to $p = 0.11$. Our calculations in the previous section
shows that as $p$ increased from 0.04 to 0.11 for the warm MHD solution,
$R_{sph}|_{wind}$ for the closest wind point decreased by a factor of 3.79.
Thus a further increase to $p \simeq 0.5$ may take the closest wind point
nearer to the black hole by a further factor of $\sim 10$ to $\sim 5 \times
10^3 r_g$. The above hypothetical numbers are assuming an almost linear change
in density as $p$ increases. In reality, the progression of the physical
quantities in the denser MHD solutions may not be that simple. We shall report
the exact calculations in our future publications. 

As our analyses stand now, even with denser warm MHD solutions with $p = 0.5$
we do not expect the wind to exist closer than $\sim 5 \times 10^3 r_g$, if
$\log \xi < 4.86$. However, note from the discussion in the previous
subsection, this distance may be reduced by a factor of $\sim 90$ to few $< 10^2
r_g$ for a modified constraint of $\log \xi < 6.0$. The density and velocity
will be increased accordingly. These speculative numbers indicate that indeed
the warm MHD outflow models may be able to explain even the most extreme winds
observed \citep{miller08, king12, diaztrigo13}. The aforementioned speculations strongly
indicate to us the kind of MHD solutions that we need to generate to fit
observations. However a confirmation of this speculations is beyond the scope
of this paper. We will report the exact calculations for the extreme MHD models
in our subsequent papers.

%%%%%%%%%%%%%%%%%%%%%%%%%%%%%%%%%%%%%%

%%%%%%%%%%%%%%%%%%%%%%%%%%%%%%%%%%%%%%
\subsection{Temperature of the outflowing gas}
\labsubsecn{subsec:TempCompare}
%%%%%%%%%%%%%%%%%%%%%%%%%%%%%%%%%%%%%%

%%%%%%%%%%%%%%%%%%%%%%%%%%%%%%%%%%%%%%%%%%%%%%%%%%%%%%%%%%%%%%%%%%%%%%%%%%%%%
\begin{figure*}
\begin{center}
\includegraphics[scale = 1.0, width = 16 cm, trim = 0 0 0 0, angle = 0]{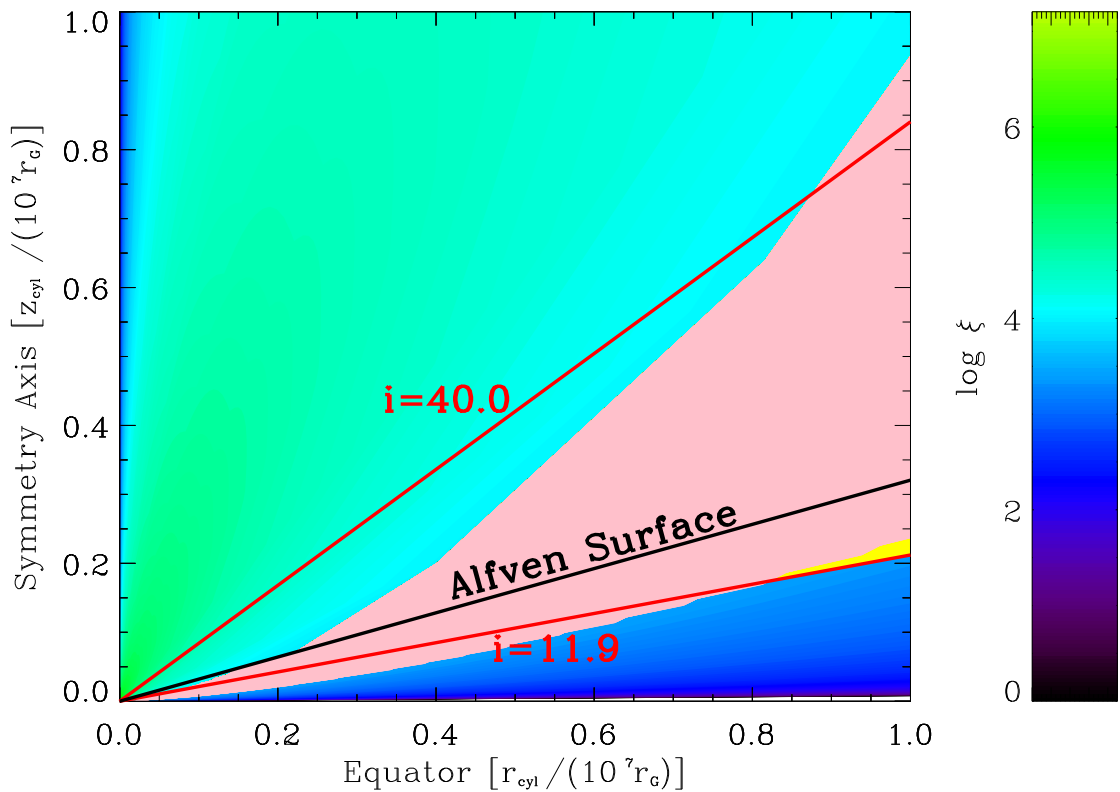}
\caption{The wind characteristics, when the ``Best Warm Solution'' is
illuminated with a Hard SED. The drastically reduced (compared to
\fig{fig:Warm}) yellow region within the outflow is obtained in the same way as
in \dfig{fig:BestCold}{fig:Warm}. We can only see a very small portion of this
yellow region at $r_{cyl}/(10^7 r_g) > 0.8$. The rest of this yellow region is
occulted by the pink wedge which represents the part of the outflow which is
thermodynamically unstable and has $3.4 \le \log \xi \le 4.05$. Note that a
small part of this unstable outflow is within the Compton thick region with
$\log N_H > 24$ (portions below the line marking the low angle $i =
11.9^{\circ}$).}  
\labfig{fig:HardInstability}
\end{center}
\end{figure*}
%%%%%%%%%%%%%%%%%%%%%%%%%%%%%%%%%%%%%%%%%%%%%%%%%%%%%%%%%%%%%%%%%%%%%%%%%%%%%

The physical properties of the MHD solutions depend on the energy equation
which involves solving the balance between the local heating and cooling
effects. Hence, along with all the other properties, like the velocity, density
etc. the temperature ($T_{MHD}$) of the outflowing gas is also specified - see
discussion in \secn{sec:ColdWarmSolns} of the Appendix. However, the MHD
calculations do not take into account the effect of photoionization of the
outflowing material due to light from the central source. In fact, the
temperature of the gas within the wind region is determined by the effects of
photoionization by the ionizing SED and may be very different from $T_{MHD}$. 

We used CLOUDY to calculate the temperature $T_{Cloudy}$ of the gas to check
how different it is from $T_{MHD}$ for the best warm solution and the
comparison is shown in \fig{fig:TempCompare}. CLOUDY calculations are
computationally expensive and hence we restricted them to within the wind
region (the shaded region with dotted red lines) of the outflow only. Note that
for the best warm solution, below $i = 11.7^{\circ}$ the gas is Compton thick
with integrated $N_H > 10^{24} cm^{-2}$. Hence photoionization and associated
acceleration of the Compton thick gas may be negligible. As such, the
properties of the outflow in the Compton thick region of the outflow is likely
to be determined by magnetic fields alone. To determine whether this
qualitative assumption is true is beyond the scope of the paper (although, see
Section 3.4 of \citealt{garcia01} for detailed methodology of how one might
attempt to solve the energy equation along a flow field line involving both the
MHD dynamical terms and a photoionization code). Hence we compare temperatures
within the Compton thin wind region only.

\fig{fig:TempCompare} shows that $T_{Cloudy}$ is indeed different from
$T_{MHD}$ and the difference increases as we move away from the surface of the
accretion disk and as we move further out. We need to judge at this point, if
this difference in the gas temperature, and hence on its enthalpy, will make a
difference to the properties of the gas. 

\dfig{fig:bern1}{fig:bern55} of the Appendix show that the specific enthalpy
term is negligible compared to the specific magnetic energy. Comparing the
specific energies, we see that even if the gas temperatures were higher (due to
photoionization) than $T_{MHD}$ by orders of magnitude, the magnetic field
would still dominate the specific energy and hence the properties of the
outflow would still be determined by the magnetic field. 

%%%%%%%%%%%%%%%%%%%%%%%%%%%%%%%%%%%%%%
\subsection{Effect of thermodynamic instability in the Hard state}
\labsubsecn{subsec:WarmHard}

Conventionally it is assumed that ionized gas cannot be detected if it is
thermodynamically unstable. \citet{chakravorty13} showed the effect of
thermodynamic considerations and found that the equilibrium curve to be
unstable for a range of $\xi$ values, but only for the Hard SED. We have
conducted stability curve analysis in \subsecn{subsec:AtomicPhysicsConstraints}
and have found similar results - for the Hard SED, the range $3.4 < \log \xi <
4.1$ is thermodynamically unstable. Thus the constraints on $\xi$ have to be
modified accordingly, when looking for the wind region within an outflow
illuminated by the Hard SED.

In \subsubsecn{subsubsec:ColdHard} we have mentioned that with the appropriate
restrictions on the $\xi$ value, no wind could be found within the cold MHD
outflows. Since the warm solutions result in much broader (than that in cold
solutions) wind region, we test if the best warm solution can have a wind with
a Hard SED.  

Using the value $\log \xi = 4.05$, we get a significant (although reduced from
the Soft SED case) wind region. Next, we check the effect of thermodynamic
instability. In \fig{fig:HardInstability} the pink region shows the part of the
outflow which has $\log \xi = 3.4 - 4.05$, a range that is ``thermodynamically
unstable''. Note that above the $i = 11.9^{\circ}$ line (which marks the
Compton thick limit), this thermodynamically unstable zone almost completely
occults the wind region (in yellow). This implies that in the Hard state, even
if a significant region of the outflow is Compton thin and has the correct
$\log \xi$ to produce FeXXVI lines, this same region is also thermodynamically
unstable. Hence in the Hard state, we cannot expect to detect the wind. 

%The results discussed in this subsection assumed that one can have warm
%solutions in the Hard state. However, warm solutions may be a characteristic
%of the Soft state only (see next \subsecn{subsec:SedWarmSoln}), making it
%further difficult for a wind detection in Hard state.   

Our analysis thus, strongly suggests that winds will not be detected in the
Hard state. Hence, we are in agreement with observational results which detect
winds only in the Softer states of the outburst \citep{ponti12}. Note that such
a correlation between accretion state (Softer) and presence of wind has been
found for neutron stars as well by \citet{ponti14}. Thus our analysis and
results maybe valid, not only for BHBs but neutron star accretion disks as
well.

We would want to discuss, at this point, an interesting observation made by
\citet{higginbottom15} on the issue of thermodynamic instability. The authors
correctly point out that, if a parcel of gas reaches a thermodynamically
unstable temperature (where the gradient of the stability curve becomes
negative - see \fig{fig:Scurve}) the gas will quickly heat up to attain the
higher temperature of the next thermodynamically stable point at same pressure
(same $\xi/T$). For a thermally driven wind this effect would result in very
efficient acceleration. However for a MHD wind this effect, particularly
relevant for the Hard State, will not aid in the wind driving mechanism. In the
Hard state (from \fig{fig:Scurve}) the maximum increase of temperature, for a
parcel of gas to avoid thermodynamic instability, is about an order of
magnitude. In the previous subsection we have demonstrated that $T_{Cloudy}$
could be much higher than $T_{MHD}$ and still not affect the properties of the
MHD driven wind. For a typical angle of $i = 20^{\circ}$, (see
\dfig{fig:TempCompare}{fig:bern55}), even if $T_{Cloudy}$ were higher by an
order of magnitude (raised by thermodynamic instability considerations), the
magnetic specific energy would still dominate over that of enthalpy. Hence, for
MHD winds, thermodynamic instability is unlikely to cause any additional
efficient acceleration. We acknowledge that thermal lifting might play a role
in the disk upper layers, where the disk material gets magneto-centrifugally
accelerated. But it is unclear whether or not photoionisation equilibrium would
correctly describe these highly expanding layers. 

%%%%%%%%%%%%%%%%%%%%%%%%%%%%%%%%%%%%%%

%%%%%%%%%%%%%%%%%%%%%%%%%%%%%%%%%%%%%%%%%%%%%%%%%%%%%%%%%%%%%%%%%%%%%%%%%%%%%

%%%%%%%%%%%%%%%%%%%%%%%%%%%%%%%%%%%%%%%%%%%%%%%%%%%%%%%%%%%%%%%%%%%%%%%%%%%%%
\section{Conclusions}
\labsecn{conclusions}

Winds are detected as absorption lines in the high resolution X-ray spectra of
black hole binaries. The absorption lines are mostly from H-like and He-like
Fe, but some rare observations show lines from other ions. \citet{ponti12} have
shown that winds are seen in the Soft state of the outburst and never in the
canonical Hard states. Further, the strongest winds were observed for objects
with high inclination angles, i.e.  the winds flow close to the disk surface at
low equatorial angles. In this paper we investigated if magneto centrifugal
outflows \citep{ferreira97, casse00b} can reproduce the observed winds in terms
of the correct range of ionization parameter ($\xi$), column density ($N_H$),
velocity ($v_{obs}$) and density $n_H$. The investigations are done as a
function of the two key accretion disk parameters - the disk aspect ratio
$\varepsilon$ and the radial exponent $p$ of the accretion rate ($\dot{M}_{acc}
\propto r^p$). We further test if our theoretical models can match the state
dependant and angle dependant nature of the accretion disk winds. The results
of our study are listed below:
\begin{enumerate}
\item[$\bullet$] The cold solutions, which are solely driven by the magnetic
acceleration, produce very narrow regions of detectable wind and from the outer
parts ($\ge 2.51 \times 10^5 r_g$) of the accretion disk. In addition, the cold
MHD winds have lower density ($\log n_H \le 9.9$) than what observations
predict. The winds were found to be equatorial, within  $i \sim 30^{\circ}$ of the
accretion disk surface.  
\item[$\bullet$] We realised that we need high values of $p (> 0.04)$ to
reproduce winds that can match observations. However $p$ cannot be increased to
desirable values in the framework of the cold MHD solutions. We definitely need
warm MHD solutions to explain the observational results. In the warm MHD
solutions, some extra heating at the disk surface causes a larger mass loading
at the base of the outflow, which is then magnetically accelerated to form a
denser wind. We speculate that the aforementioned heating may be due to the
illuminating SED, particularly in the Soft state, or due to dissipation of
energy by MHD turbulence, within the disk. Indeed, even a few percents of the
released accretion energy (if it were deposited on the disk surface, leading to
local heating there, instead of being dissipated deep within the disk layers),
would be enough to give rise to such warm MHD winds. Whether or not MHD
turbulence in accretion disks provides such a surface heating is an open
theoretical issue. 
%But observations of winds from BHB could possibly lead to interesting constraints. 
\item[$\bullet$] In the Soft state, our densest warm MHD solution predicts a
wind at $7.05 \times 10^4 r_g$ with a density of $\log n_H = 11.1$. The densest
part of the wind ($\log n_H > 8$) still remains equatorial - within $i \sim
30^{\circ}$ of the accretion disk. The values of the physical parameters are
consistent with some of the observed winds in BHBs. However, there are some
other extreme observations \citep[e.g of GRO J1655-40][]{miller08} which
require a denser wind which is at a smaller distance to the black hole. From
our work we understand what kind of MHD solutions can reproduce such extreme
winds - warm MHD solutions with $p \simeq 0.5$. It was beyond the scope of this
paper to produce those particular solutions. However, we will generate and
report such solutions in our future publications where we will attempt to
reproduce spectra of BHB winds of different kinds.  
\item[$\bullet$] The outflow illuminated by a Hard SED will not produce
detectable wind because (i) the allowed region of the winds is smaller
(compared to the Soft SED case) and (b) the wind region falls within the
thermodynamically unstable range of $\log \xi$ and hence unlikely to be
detected.   
\item[$\bullet$] Thus in the framework of MHD outflows we can satisfy the
observed trends reported in \citet[][and references therein]{ponti12} that -
(a) winds are observed in the Soft states (and not expected in the Hard state)
of the BHB outbursts and (b) accretion disk winds in BHBs are equatorial. We
have been able to reproduce the expected values (consistent with observations)
of distance, density and velocity for the average winds in BHBs. For the
extremely dense (and hence at small distances from the black hole) winds our
rigorous analysis was capable of pointing to the kind of accretion disks which
will be able to reproduce them.  
\end{enumerate} 

%%%%%%%%%%%%%%%%%%%%%%%%%%%%%%%%%%%%%%%%%%%%%%%%%%%%%%%%%%%%%%%%%%%%%%%%%%%%%%

%%%%%%%%%%%%%%%%%%%%%%%%%%%%%%%%%%%%%%%%%%%%%%%%%%%%%%%%%%%%%%%%%%%%%%%%%%%%%%
%\begin{acknowledgments}
%\section{acknowledgments}
\begin{acknowledgements}
The authors acknowledge funding support from the French Research National
Agency (CHAOS project ANR-12-BS05-0009 http://www.chaos-project.fr) and CNES.
\end{acknowledgements}
%\end{acknowledgments}

%%%%%%%%%%%%%%%%%%%%%%%%%%%%%%%%%%%%%%%%%%%%%%%%%%%%%%%%%%%%%%%%%%%%%%%%%%%%%%

%%%%%%%%%%%%%%%%%%%%%%%%%%%%%%%%%%%%%%%%%%%%%%%%%%%%%%%%%%%%%%%%%%%%%%%%%%%%%%

\appendix

\section{Cold versus warm disk wind solutions}
\labsecn{sec:ColdWarmSolns}

%%%%%%%%%%%%%%%%%%%%%%%%
\begin{figure}
\begin{center}
\includegraphics[height = 0.9\columnwidth, trim = 0 50 30 240, clip, angle = -90]{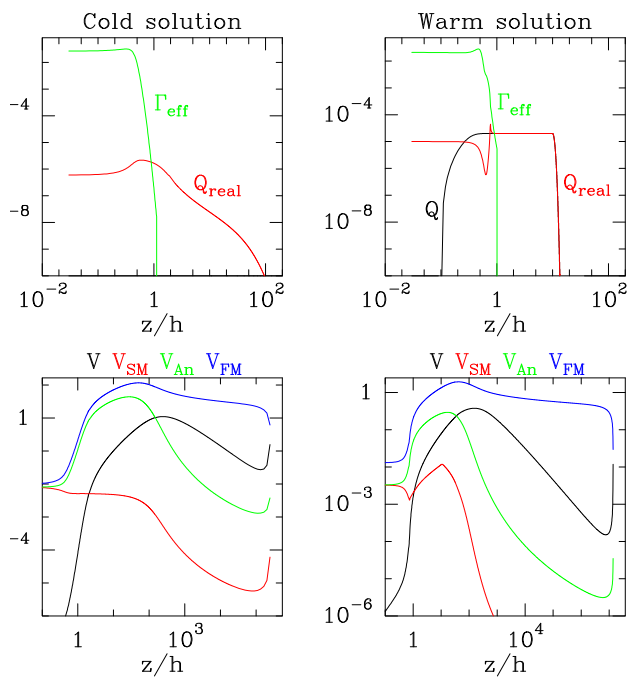}
\caption{Profiles along a magnetic surface of a typical cold -isothermal-
solution with $p=0.006$ (left) and warm solution with $p=0.1$ (right). Top:
effective turbulent heating $\Gamma_{eff}$, real entropy generation term
$Q_{real}$ and prescribed function $Q$ in arbitrary units. Bottom: critical
velocities (see text) normalized to the disk mid plane sound speed.}
\labfig{fig:2sol_Q}
\end{center}
\end{figure}
%%%%%%%%%%%%%%%%%%%%%%%%

In this appendix, we highlight some important points allowing to distinguish
between "cold" and "warm" wind solutions from near-Keplerian accretion disks.
In the terminology of \cite{blandford82}, a cold wind refers to a flow where
the enthalpy is negligible with respect to the magnetic energy density at the
base, namely at the disk surface. Unless some additional heating source occurs
within the disk layers and/or at its surface, the temperature of the flow
leaving the disk is at most comparable to that prevailing at the disk mid
plane. This translates into an enthalpy which is roughly $(h/r)^2$ times the
gravitational potential, hence negligible in a thin disk. As a consequence, a
wind with a positive Bernoulli integral can only be achieved by magnetic means.
Cold models have been thus computed using different prescriptions for the
thermal state of the magnetic surfaces: either isothermal (constant temperature
along a surface, eg. \cite{ferreira97}) or adiabatic surfaces (decreasing
temperature, eg \cite{casse00a}). 

%
%%%%%%%%%%%%%%%%%%%%%%%%
\begin{figure}
\begin{center}
\includegraphics[height = 0.9\columnwidth, trim = 0 30 30 240, clip, angle = -90]{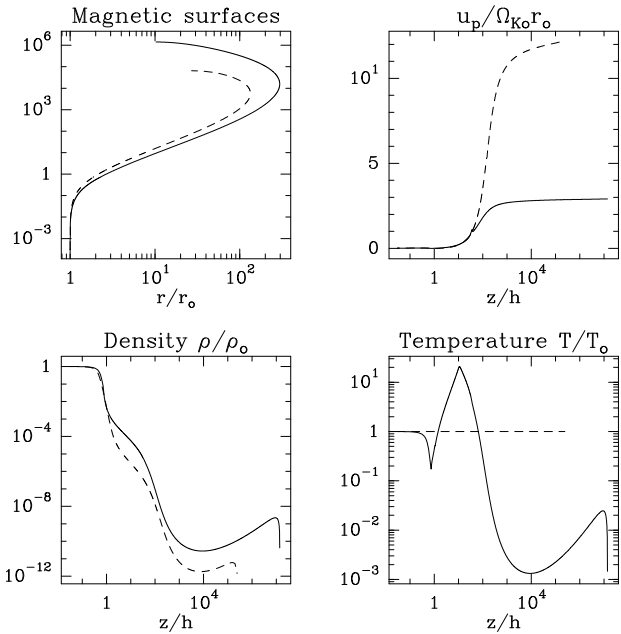}
\caption{Shape of the magnetic surfaces (top left) and vertical profiles along
the magnetic surfaces of (a) outflow poloidal velocity normalized to the
Keplerian speed at the field line anchoring radius $r_o$ (top right), (b)
density normalized to its mid plane value (bottom left) and (c) temperature
normalized to its mid plane value (bottom right). Solid lines are for the warm
solution, dashed for the isothermal.}  
\labfig{fig:2sol_prof}
\end{center}
\end{figure}
%%%%%%%%%%%%%%%%%%%%%%%%

On the contrary, warm disk wind models rely on the existence of some ad-hoc
entropy generation term $Q$. The exact energy equation for the outflowing
material writes 
\begin{equation}
\rho T\frac{dS}{dt} =  \rho T \vec{u_p} \cdot \vec{ \nabla} S = Q_{real}
\label{eq:Qreal}
\end{equation}  
\noindent where $S$ is the specific entropy and $Q_{real}$ is the local source
of entropy which arises from the difference between all heating and cooling
processes  (\cite{casse00b}). More specifically, it can be written $Q_{real} =
(\Gamma_{eff} + \Gamma_{turb} + \Gamma_{ext}) - (\Lambda_{rad} +
\Lambda_{turb})$, where $\Gamma_{eff} = \eta_mJ^2_{\phi} + \eta'_mJ^2_p +
\eta_v r\left|\vec{ \nabla}\Omega\right|^2$ is the effective Joule and viscous
heating, $\Gamma_{turb}$ is a turbulent heating term that cannot be described
by simple anomalous transport coefficients (namely the term $\Gamma_{eff}$) and
would correspond, for instance, to some resonant or wave heating above the
disk, $\Gamma_{ext}$ is an external source of energy (typically due to some
illumination by UV or X-rays, if present at all). 
$\Lambda_{rad}= \vec{\nabla} \cdot \vec S_{rad}$ is the radiative cooling
($\vec S_{rad}$ being the radiative flux) and $\Lambda_{turb}$ is a cooling due
to a turbulent energy transport, which is most probably also taking place
inside turbulent disks (see \cite{casse00b} for more details).

Taking into account all these processes is a tremendous task which requires the
understanding not only of MHD turbulence in outflow emitting disks, but also of
the complex radiative processes at work in various astrophysical objects. A
simplified approach is actually to assume the entropy generation term, which is
what we do. The above exact energy equation follows the flow streamlines, which
is inconvenient to use when integrating the equations from the disk mid plane
to outflow asymptotics (isothermal solutions cannot be obtained for instance).
Hence, instead, the energy equation actually solved is 
\begin{equation}
\frac{\rho u_p}{B_p}  \vec{B_p} \cdot \nabla C_s^2 = (\gamma-1) \left ( Q + C_s^2  \frac{\rho u_p}{B_p}  \vec{B_p} \cdot \nabla \ln \rho \right )
\end{equation}  
where $\gamma$ is the adiabatic index and $C_s^2= P/\rho$ with a prescribed
self-similar function $Q$ described in Section 4.1 of \cite{casse00b}. This
equation is strictly equivalent to Eq.(\ref{eq:Qreal}) in the ideal MHD outflow
region and allows to treat the disk vertical structure. Once a full
trans-Alfv\'enic MHD flow solution is obtained, the real entropy generation
term $Q_{real}$ can be computed using Eq.(\ref{eq:Qreal}) as shown in
\fig{fig:2sol_Q}.

To illustrate cold and warm solutions, we choose two representative
super-Alfvenic solutions with same $\varepsilon=0.01$. The cold solution is
isothermal ($\gamma=1$) with $p=0.006$, whereas the warm solution with
$\gamma=5/3$ has $p=0.1$ and requires a heating function $Q$.  
The upper panels of \fig{fig:2sol_Q} show, for both solutions, the vertical
profiles along a magnetic surface of the effective turbulent heating
$\Gamma_{eff}$ and the imposed $Q$, as well as the real $Q_{real}$, entropy
generation terms. The entropy parameter, defined as  
\begin{equation}
f = \frac{ \int_{V} Q_{real} dV}{\int_{disk} \Gamma_{eff} dV}  
\end{equation}
where the volume $V$ is both the disk and the wind, can be computed a
posteriori once a solution is found. It provides the ratio of the power due to
the extra heating going into the wind to the turbulent power dissipated within
the disk. The warm solution resulted in an entropy parameter f = 0.02, which
suggests that local MHD turbulence could actually lead to such solution.
Indeed, the required extra heating amounts only to 2\% of the power that would
be dissipated within the disk (hence a reduction of 2\% of the disk
luminosity). This would be possible if MHD turbulence itself ($(\Gamma
-\Lambda)_{turb}$ terms) conveys that power to the disk upper layers (the ’base
of the wind’). This is an open theoretical issue of course. If such a process
proves to be inexistent, then one should rely only on illumination
($\Gamma_{ext}$ term) to obtain warm MHD solutions of this kind.  
Note that magneto-centrifugal winds undergo a huge adiabatic cooling at the
disk surface so that to remain isothermal requires some heat deposition as
well. Thus, the cold solution displayed here would require a $Q_{real}$ such
that $f=1.6 \times 10^{-3}$. The lower panels show the profiles of the various
velocities relevant in such MHD flows: the critical flow speed $V$, the slow
$V_{SM}$ and fast $V_{FM}$ magnetosonic phase speeds and the Alfven speed
$V_{An}$ (see \citealt{ferreira95} for their meaning and definition). Note that
the warm solution becomes super-SM above the disk but that  $V_{SM}$ is always
smaller than the local sound speed.

The resulting solutions are shown in \fig{fig:2sol_prof}. While the location of
the SM point remains in both cases roughly above the disk surface
($x_{SM}=z/h=2.08$ in the cold case,  $x_{SM}=1.1$ in the warm case), the main
difference introduced by the surface heating term is the existence of a radial
pressure gradient above the disk surface enforcing the wind to open up
(\cite{ferreira04}). Thus, while the Alfven surface for the cold solution is
located at $x_A=151.5$, namely $z_A/r_o= 14.61$, $r_A/r_o= 9.64$ or an angle
$\Phi_A=33^o$ from the vertical axis, it is much closer to the disk surface in
the warm case, with $x_A=36.2$, namely $z_A/r_o= 0.92$, $r_A/r_o= 2.54$ or
$\Phi_A=70^o$.    
The overall outflow behavior remains however the same: after an initial
widening up to a maximum distance, the flow undergoes a recollimation towards
the jet axis (perpendicular to the disk) where an oblique shock is expected to
occur (and the validity of the self-similar solution breaks down,
\cite{ferreira97}). Due to the heating term present in the warm solution, the
flow temperature is seen to increase up to about 20 times the mid plane
temperature before undergoing an adiabatic decrease once the heating vanishes
($Q=0$). Note that the temperature profile mostly affects the disk vertical
balance, allowing thereby a larger mass loss (larger ejection index $p$) in the
warm case than in the isothermal case. But the asymptotic outflow speed is
mainly a result of the magnetic energy (dominant term) conversion. This is
illustrated in \fig{fig:bern1} and \fig{fig:bern55} which show, for the cold
and warm cases respectively, the profiles of the various specific energy
reservoirs along a magnetic surface (the sum of which defines the Bernoulli
invariant). In self-similar solutions of this kind, all the initial magnetic
energy is eventually converted into outflow kinetic poloidal energy.

%%%%%%%%%%%%%%%%%%%%%%%%
\begin{figure}
\begin{center}
\includegraphics[width = 0.9\columnwidth, trim = 10 10 110 75, clip ]{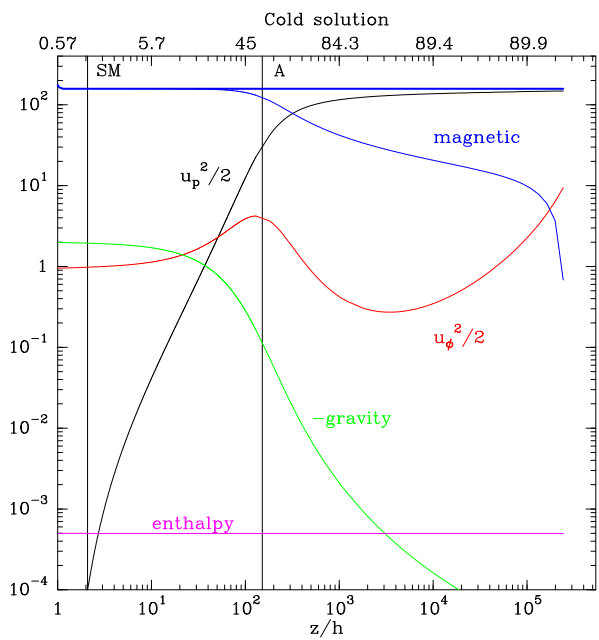}
\caption{Profiles of various specific energy reservoirs (in $GM/2r_o$ units)
along a magnetic surface anchored at a radius $r_o$, for the cold isothermal
solution with $p=0.006$. The SM-point is located at $x_{SM}=z/h=2.08$ and the
Alfven point at $x_A=151.5$. The thick solid blue line is the sum of all terms
and defines the Bernoulli invariant. At the top of the figure, the axis is
labelled with the inclination angle (in degrees) from the disk midplane $i=
atan(\varepsilon z/h)$.}  
\labfig{fig:bern1}
\end{center}
\end{figure}
%%%%%%%%%%%%%%%%%%%%%%%%
%%%%%%%%%%%%%%%%%%%%%%%%
\begin{figure}
\begin{center}
\includegraphics[width = 0.9\columnwidth, trim = 10 10 110 75, clip ]{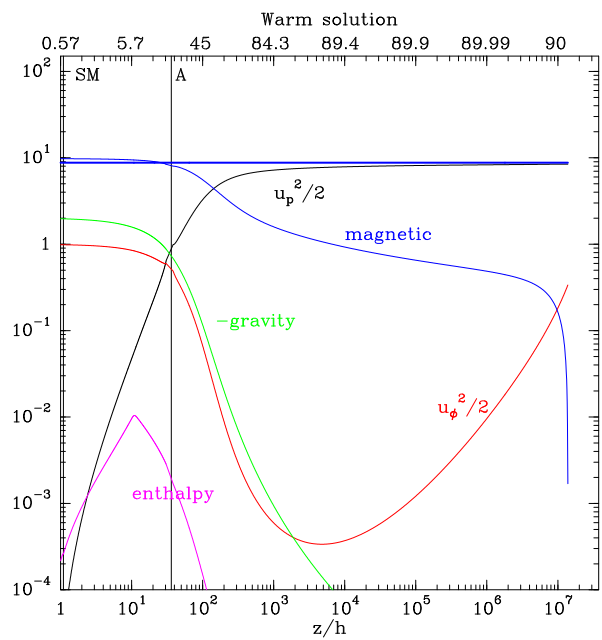}
\caption{Same as \fig{fig:bern1}, but for the best warm ($\varepsilon = 0.01,
\,\, p=0.1$) solution. The SM-point is located at $x_{SM}=1.1$ and the Alfven
point at $x_A=36.2$. Note the drastic decrease of the enthalpy due to the
adiabatic cooling, once the heating term vanishes.}   
\labfig{fig:bern55}
\end{center}
\end{figure}
%%%%%%%%%%%%%%%%%%%%%%%%

%%%%%%%%%%%%%%%%%%%%%%%%%%%%%%%%%%%%%%%%%%%%%%%%%%%%%%%%%%%%%%%%%%%%%%%%%%%%%%
\bibliographystyle{aa}

\begin{thebibliography}{}
%
%\bibitem[\protect\citeauthoryear{Allende Prieto \etal}{2001}]{allendeprieto01} Allende Prieto, C., Lambert, D.L., \& Asplund, M., 2001, ApJ, 556, L63

%\bibitem[\protect\citeauthoryear{Allende Prieto \etal}{2002}]{allendeprieto02} Allende Prieto, C., Lambert, D.L., \& Asplund, M., 2002, ApJ, 573, L137

\bibitem[\protect\citeauthoryear{Arnaud}{1996}]{arnaud96} Arnaud, K. A. 1996, ASPC, 101, 17	

\bibitem[\protect\citeauthoryear{Blandford \& Payne}{1982}]{blandford82} Blandford, R. D.; Payne, D. G. 1982, MNRAS, 199, 883

\bibitem[\protect\citeauthoryear{Blum \etal}{2010}]{blum10} Blum, J. L.; Miller, J. M.; Cackett, E.; Yamaoka, K.; Takahashi, H.; Raymond, J.; Reynolds, C. S.; Fabian, A. C. 2010, ApJ, 713, 1244

%\bibitem[\protect\citeauthoryear{Brandt \& Schulz}{2000}]{brandt00} Brandt, W. N.; Schulz, N. S. 2000, ApJ, 544L, 123

\bibitem[\protect\citeauthoryear{Casse \& Ferreira}{2000a}]{casse00a} Casse, F.; Ferreira, J. 2000, A\&A, 353, 1115

\bibitem[\protect\citeauthoryear{Casse \& Ferreira}{2000b}]{casse00b} Casse, F.; Ferreira, J. 2000, A\&A, 361, 1178

%\bibitem[\protect\citeauthoryear{Canizares \etal}{2005}]{canizares05} Canizares, C. \etal 2005, PASP, 117, 1144

\bibitem[\protect\citeauthoryear{Chakravorty \etal}{2008}]{chakravorty08} Chakravorty, S., Kembhavi, A.K., Elvis, M. \& Ferland, G., Badnell, N. R. 2008, MNRAS, 384L, 24

\bibitem[\protect\citeauthoryear{Chakravorty \etal}{2009}]{chakravorty09} Chakravorty, S., Kembhavi, A.K., Elvis, M. \& Ferland, G., 2009, MNRAS, 393, 83
%
\bibitem[\protect\citeauthoryear{Chakravorty \etal}{2012}]{chakravorty12} Chakravorty, S., Misra, R., Elvis, M., Kembhavi, A.K., \& Ferland, G., 2009, MNRAS, 393, 83

\bibitem[\protect\citeauthoryear{Chakravorty \etal}{2013}]{chakravorty13} Chakravorty, S., Lee, J. C., Neilsen, J. 2013, MNRAS, 436, 560
%
\bibitem[\protect\citeauthoryear{Contopoulos \& Lovelace}{1994}]{contopoulos94} Contopoulos, J., \& Lovelace, R. V. E. 1994, ApJ, 429, 139

\bibitem[\protect\citeauthoryear{Diaz Trigo \etal}{2013}]{diaztrigo13} Diaz Trigo, M; Miller-Jones, J.C.A.; Migliari, S; Broderick, J.W.; Tzioumis, T; 2013, Nature, 504, 260 

\bibitem[\protect\citeauthoryear{Ferland \etal}{1998}]{ferland98} Ferland, G. J.; Korista, K. T.; Verner, D. A.; Ferguson, J. W.; Kingdon, J. B.; Verner, E. M. 1998, PASP, 110, 761

\bibitem[\protect\citeauthoryear{Ferreira}{1993}]{ferreira93} Ferreira, J., \& Pelletier, G. 1993, A\&A, 276, 625

\bibitem[\protect\citeauthoryear{Ferreira}{1995}]{ferreira95} Ferreira, J.; Pelletier, G. 1995, A\&A, 295, 807

\bibitem[\protect\citeauthoryear{Ferreira}{1997}]{ferreira97} Ferreira, J. 1997, A\&A, 319, 340

\bibitem[\protect\citeauthoryear{Ferreira}{2004}]{ferreira04} Ferreira, J.; Casse, F. 2004, ApJ, 601L, 139

\bibitem[\protect\citeauthoryear{Ferreira \etal}{2006}]{ferreira06} Ferreira, J.; Petrucci, P.-O.; Henri, G.; Saugé, L.; Pelletier, G. 2006, A\&A, 447, 813

\bibitem[\protect\citeauthoryear{Frank, King \& Raine}{2002}]{frank02} Frank, J., King, A., \& Raine, D. 2002, Accretion Power in Astrophysics (3rd ed.; Cambridge: Cambridge Univ. Press)

\bibitem[\protect\citeauthoryear{Fukumura \etal}{2010a}]{fukumura10a} Fukumura, K.; Kazanas, D.; Contopoulos, I.; Behar, E. 2010, ApJ, 715, 636	

\bibitem[\protect\citeauthoryear{Fukumura \etal}{2010b}]{fukumura10b} Fukumura, K.; Kazanas, D.; Contopoulos, I.; Behar, E. 2010, ApJ, 723L, 228	

\bibitem[\protect\citeauthoryear{Fukumura \etal}{2014}]{fukumura14} Fukumura, K.; Tombesi, F.; Kazanas, D.; Shrader, C.; Behar, E.; Contopoulos, I.  2014, ApJ, 780, 120

\bibitem[\protect\citeauthoryear{Fukumura \etal}{2015}]{fukumura15} Fukumura, K.; Tombesi, F.; Kazanas, D.; Shrader, C.; Behar, E.; Contopoulos, I. 2015, ApJ, 805, 17

\bibitem[\protect\citeauthoryear{Garcia \etal}{2001}]{garcia01} Garcia, P. J. V.; Ferreira, J.; Cabrit, S.; Binette, L. 2001, A\&A, 377, 589

%\bibitem[\protect\citeauthoryear{Greiner \etal}{2001}]{greiner01} Greiner, J.; Cuby, J. G.; McCaughrean, M. J. 2001, Nature, 414, 522

%\bibitem[\protect\citeauthoryear{Grevesse \& Sauval}{1998}]{grevesse98} Grevesse, N., \& Sauval, A.J., 1998, Space Science Review, 85, 161

% \bibitem[\protect\citeauthoryear{Hanke \etal}{2009}]{hanke09} Hanke, M.; Wilms, J.; Nowak, M. A.; Pottschmidt, K.; Schulz, N. S.; Lee, J. C. 2009, ApJ, 690, 330

% \bibitem[\protect\citeauthoryear{Hess \etal}{1997}]{hess97} Hess, C.J.; Kahn, S.M.; Paerels, F.B.S., 1997, ApJ, 478, 94

\bibitem[\protect\citeauthoryear{Higginbottom \& Proga}{2015}]{higginbottom15} Higginbottom, N.; Proga, D. 2015, ApJ, 807, 107

%\bibitem[\protect\citeauthoryear{Holweger}{2001}]{holweger01} Holweger, H., 2001, Joint SOHO/ACE workshop ``Solar and Galactic Composition''. Edited by Robert F. Wimmer-Schweingruber. Publisher: American Institute of Physics Conference proceedings, 598, 23

%\bibitem[\protect\citeauthoryear{Jimenez-Garate \etal}{2002}]{jimenez-garate02} Jimenez-Garate, M. A., Raymond, J. C., \& Liedahl, D. A. 2002, ApJ, 581, 1297

\bibitem[\protect\citeauthoryear{Kallman \etal}{2009}]{kallman09} Kallman, T. R.; Bautista, M. A.; Goriely, S.; Mendoza, C.; Miller, J. M.; Palmeri, P.; Quinet, P.; Raymond, J. 2009, ApJ, 701, 865

\bibitem[\protect\citeauthoryear{King \etal}{2012}]{king12} King, A. L. \etal 2012, ApJ, 746L, 20

%\bibitem[\protect\citeauthoryear{Krolik \etal}{Krolik \etal}{1981}]{krolik81} Krolik, J. H., McKee, C. F., \& Tarter, C. B. 1981, ApJ, 249, 422

\bibitem[\protect\citeauthoryear{Kubota \etal}{2007}]{kubota07} Kubota \etal 2007, PASJ, 59S, 185

\bibitem[\protect\citeauthoryear{Lee \etal}{2002}]{lee02} {Lee}, J.~C. and {Reynolds}, C.~S. and {Remillard}, R. and {Schulz}, N.~S. and {Blackman}, E.~G. and {Fabian}, A.~C. 2002, \apj, 567, 1102

%\bibitem[\protect\citeauthoryear{Lee \etal}{2013}]{lee13} Lee, J. C \etal 2013, In Press, arXiv1301.3148

\bibitem[\protect\citeauthoryear{Mitsuda \etal}{1984}]{mitsuda84} Mitsuda, K. \etal 1984, PASJ, 36, 741

\bibitem[\protect\citeauthoryear{Makishima \etal}{1987}]{makishima86} Makishima, K.; Maejima, Y.; Mitsuda, K.; Bradt, H. V.; Remillard, R. A.; Tuohy, I. R.; Hoshi, R.; Nakagawa, M. 1986, ApJ, 308, 635
%
\bibitem[\protect\citeauthoryear{Miller \etal}{2004}]{miller04} Miller \etal 2004, ApJ, 601, 450 

\bibitem[\protect\citeauthoryear{Miller \etal}{2006}]{miller06}  Miller, J. M.; Raymond, J.; Homan, J.; Fabian, A. C.; Steeghs, D.; Wijnands, R.; Rupen, M.; Charles, P.; van der Klis, M.; Lewin, W. H. G. 2006, ApJ, 646, 394

\bibitem[\protect\citeauthoryear{Miller \etal}{2008}]{miller08} {Miller}, J.~M. and {Raymond}, J and {Reynolds}, C.~S. and {Fabian}, A.~C. and {Kallman}, T.~R. and {Homan}, J. 2008, \apj, 680, 1359

%\bibitem[\protect\citeauthoryear{Miller \etal}{2011}]{miller11} Miller, Jon M.; Maitra, Dipankar; Cackett, Edward M.; Bhattacharyya, Sudip; Strohmayer, Tod E. 2011, ApJ, 731L, 7

\bibitem[\protect\citeauthoryear{Miller \etal}{2012}]{miller12} Miller \etal 2012, ApJ, 759L, 6

\bibitem[\protect\citeauthoryear{Morrison \& McCammon}{1983}]{morrison83} Morrison, R.; McCammon, D. 1983, ApJ 270, 119

\bibitem[\protect\citeauthoryear{Neilsen \& Lee}{2009}]{neilsen09} Neilsen, J; Lee, J. C. 2009, Natur, 458, 481

%\bibitem[\protect\citeauthoryear{Neilsen \etal}{2009}]{neilsen09b} Neilsen, J; Lee, J. C.; Nowak, M. A.; Dennerl, K.; Vrtilek, S. D. 2009, ApJ, 696, 182

\bibitem[\protect\citeauthoryear{Neilsen \etal}{2011}]{neilsen11} Neilsen, J.; Remillard, R. A.; Lee, J. C. 2011, ApJ, 737, 69

\bibitem[\protect\citeauthoryear{Neilsen \& Homan}{2012}]{neilsen12a} Neilsen, J.; Homan, J. arXiv1202.6053

%\bibitem[\protect\citeauthoryear{Neilsen \etal}{2012}]{neilsen12b} Neilsen, J.; Petschek, A. J.; Lee, J. C. 2012, MNRAS, 421, 502

%\bibitem[\protect\citeauthoryear{Orosz \etal}{1997}]{orosz97} Orosz, J., \& Baily, C. D. 1997, ApJ, 477, 876

%\bibitem[\protect\citeauthoryear{Orosz \etal}{2004}]{orosz04} Orosz, J.; McClintock, J. E.; Remillard, R. A.; Corbel, S. 2004, ApJ, 616, 376

%\bibitem[\protect\citeauthoryear{Orosz \etal}{2007}]{orosz07} Orosz, J.\etal 2007, Nature, 449, 872O,

%\bibitem[\protect\citeauthoryear{Orosz \etal}{2011}]{orosz11} Orosz, J.; McClintock, J. E.; Aufdenberg, J. P.; Remillard, R. A.; Reid, M. J.; Narayan, R.; Gou, L. 2011, ApJ, 742, 84

\bibitem[\protect\citeauthoryear{Peterson}{1997}]{peterson97} Peterson, B. M. 1997, An Introduction to Active Galactic Nuclei. Cambridge Univ. Press, Cambridge.
%

%\bibitem[\protect\citeauthoryear{Petrucci \etal}{2008}]{petrucci08} Petrucci, Pierre-Olivier; Ferreira, Jonathan; Henri, Gilles; Pelletier, Guy, 2008, MNRAS, 385L 88

\bibitem[\protect\citeauthoryear{Petrucci \etal}{2010}]{petrucci10} Petrucci, Pierre-Olivier; Ferreira, Jonathan; Henri, Gilles; Malzac, J.; Foellmi, C. 2010, A\&A, 522, 38

\bibitem[\protect\citeauthoryear{Ponti \etal}{2014}]{ponti14} Ponti, G.; Munoz-Darias, T; Fender, R.P. 2014, MNRAS, 444, 1829

\bibitem[\protect\citeauthoryear{Ponti \etal}{2012}]{ponti12} Ponti, G.; Fender, R. P.; Begelman, M. C.; Dunn, R. J. H.; Neilsen, J.; Coriat, M. 2012, MNRAS, 422L, 11	

%\bibitem[\protect\citeauthoryear{Proga et.al}{1998}]{proga98} Proga, D.; Stone, J. M.; Drew, J. E. 1998, MNRAS, 295, 595	

%\bibitem[\protect\citeauthoryear{Proga}{2003}]{proga03} Proga, D. 2003, ApJ, 585, 406	

\bibitem[\protect\citeauthoryear{Remillard \& McClintock}{2006}]{remillard06} {Remillard}, R.~A. and {McClintock}, J.~E. 2006,  Annu. Rev. Astron. Astrophys. 44, 49 

%\bibitem[\protect\citeauthoryear{Reynolds \& Fabian}{1995}]{reynolds95} Reynolds, C.S. \& Fabian, A.C., 1995, MNRAS, 273, 1167

%\bibitem[\protect\citeauthoryear{Reynolds \& Miller}{2010}]{reynolds10} Reynolds, M. T.; Miller, J. M. 2010, ApJ, 723, 1799

%\bibitem[\protect\citeauthoryear{Reynolds}{2012}]{reynolds12} Reynolds, C.S. 2012, ApJ, 759L, 15

\bibitem[\protect\citeauthoryear{Schulz \& Brandt}{2002}]{schulz02} Schulz, N. S.; Brandt, W. N. 2002, ApJ, 572, 971

\bibitem[\protect\citeauthoryear{Tarter \etal}{1969}]{tarter69} Tarter, C.B., Tucker, W. \& Salpeter, E.E., 1969, ApJ 156, 943

\bibitem[\protect\citeauthoryear{Ueda \etal}{2004}]{ueda04} Ueda, Y.; Murakami, H.; Yamaoka, K.; Dotani, T.; Ebisawa, K. 2004, ApJ, 609, 325

\bibitem[\protect\citeauthoryear{Ueda \etal}{2009}]{ueda09} {Ueda}, Y. and {Yamaoka}, K. and {Remillard} R.~A. 2009, \apj, 695, 888.

\bibitem[\protect\citeauthoryear{Ueda \etal}{2010}]{ueda10} {Ueda}, Y. \etal 2010, ApJ, 713, 257

%\bibitem[\protect\citeauthoryear{Zimmerman}{2005}]{zimmerman05} Zimmerman, E. R.; Narayan, R.; McClintock, J. E.; Miller, J. M. 2005, ApJ, 618, 832

\end{thebibliography}

%%%%%%%%%%%%%%%%%%%%%%%%%%%%%%%%%%%%%%%%%%%%%%%%%%%%%%%%%%%%%%%%%%%%%%%%%%%%%%

\end{document}